\documentclass[traditabstract]{aa}  
\usepackage{txfonts}
\usepackage{epstopdf}
\usepackage[switch]{lineno}
\usepackage{graphicx,longtable,lscape,natbib,amssymb,amssymb,amsmath}
\bibpunct{(}{)}{;}{a}{}{,} 
\newcommand{\Ha} {H$\alpha$}  
\newcommand{\Hb} {H$\beta$}  
\newcommand{\ergsHz}{\>{\rm erg}\,{\rm s}^{-1}\,{\rm Hz}^{-1}}
\newcommand{\kms}{$\rm{\,km \,s}^{-1}$}
\usepackage{color}

\begin{document}

\title{The MURALES survey} \subtitle{VII. Optical spectral properties
  of the nuclei of 3C radio sources at $0.3 < z < 0.82$.}

\author{Alessandro Capetti\inst{1} \and Barbara Balmaverde\inst{1}
  \and R.D. Baldi\inst{2} \and S. Baum\inst{3} \and
  M. Chiaberge\inst{4,5} \and P. Grandi\inst{6} \and
  A. Marconi\inst{7,8} \and C. O$'$Dea\inst{3} \and
  G. Venturi\inst{9,8} }

\institute { INAF - Osservatorio Astrofisico di Torino, Via
  Osservatorio 20, I-10025 Pino Torinese, Italy \and INAF- Istituto di
  Radioastronomia, Via Gobetti 101, I-40129 Bologna, Italy \and
  Department of Physics and Astronomy, University of Manitoba,
  Winnipeg, MB R3T 2N2, Canada \and Space Telescope Science Institute,
  3700 San Martin Dr., Baltimore, MD 21210, USA \and Johns Hopkins
  University, 3400 N. Charles Street, Baltimore, MD 21218, USA \and
  INAF - Osservatorio di Astrofisica e Scienza dello Spazio di
  Bologna, via Gobetti 93/3, 40129 Bologna, Italy \and Dipartimento di
  Fisica e Astronomia, Universit\`a di Firenze, via G. Sansone 1,
  50019 Sesto Fiorentino (Firenze), Italy \and INAF - Osservatorio
  Astrofisico di Arcetri, Largo Enrico Fermi 5, I-50125 Firenze,Italy
  \and Instituto de Astrof\'isica, Facultad de
  F\'isica, Pontificia Universidad Cat\'olica de Chile, Casilla 306,
  Santiago 22, Chile }

\date{} 

\abstract{This paper is the seventh work in the MUse RAdio Loud Emission lines
  Snapshot (MURALES) project series, presenting the results of observations
  obtained with the VLT/MUSE integral field spectrograph of 3C radio
  sources. Here, we discuss the optical spectral properties of the nuclei
  of 26 objects with $0.3 < z < 0.82$ (median redshift 0.51). At
  these redshifts, the \Ha\ and [N~II] emission lines are not covered
  by optical spectra and alternative diagnostic diagrams are needed to
  separate the different spectroscopic sub-classes. We derived a robust
  spectroscopic classification into high and low-excitation galaxies
  (HEGs and LEGs) by only using the ratios of emission lines in the rest-frame UV and the blue portion of the spectra. A key result  of this study is that
  FR~II  LEGs are also found  at the highest level of radio power (up to
  $L_{178 \rm{MHz}} \sim 2\times10^{35} \ergsHz$), placing them among the most
  luminous radio sources in the Universe. Furthermore, their fraction
  within the FR~II RG population does not strongly depend on radio
  luminosity. This suggests that the jet properties in powerful
    FR~II radio sources do not depend on the accretion mode or on
  the structure of the accretion disk -- as would otherwise be expected if the jet launching
  process were due to the extraction of the rotational energy of the
  supermassive black hole. The alternative possibility of recurrent
  transitions between a LEG and a HEG phase is disfavored based on the
  variation timescales of the various active galactic nucleus (AGN) components.}

\titlerunning{The MURALES survey VII: nuclear properties of 3C radio
  sources at $0.3 < z < 0.82$} \authorrunning{A. Capetti et al.}
\maketitle

\section{Introduction}
\label{intro}

Optical spectroscopic information plays a major role in improving our  understanding of the properties of the central engines of
active galactic nuclei (AGNs). \citet{heckman80} and \citet{baldwin81}
proposed the use of optical line ratios as diagnostic tools to classify
emission-line objects in general -- and AGNs in particular. They
introduced diagnostic diagrams (DDs) comparing selected emission line
ratios showing that these are able to distinguish H~II regions ionized
by young stars from gas clouds ionized by nuclear activity. Moreover,
AGNs were separated into Seyferts and Low Ionization Nuclear
Emission-line Regions \citep[LINERs,][]{heckman80} based on the
relative ratios of the optical oxygen lines ([O I]$\lambda$6364, [O
  II]$\lambda$3727, and [O III]$\lambda$5007). Subsequently
\citet{veilleux87} revised the definition of the DDs, using only
ratios of lines with small separation in wavelength, thus removing the
problems related to reddening as well as to uncertainties on the flux
calibration of the spectra. These authors used the following line combinations in their study:
[O~III]/H$\beta$ as a function of [N~II]$\lambda$6584/\Ha,
[S~II]$\lambda\lambda$6716,6731/\Ha, and [O~I]/\Ha. \citet{kewley06}
selected emission line galaxies from the SDSS, finding that Seyferts
and LINERs form separated branches in the DDs. They suggested that the
observed dichotomy corresponds to the presence of two sub-populations
of AGNs associated with different accretion states.

An attempt to adopt a similar scheme for the optical classification
focusing on radio-galaxies (RGs) was made by \citet{laing94}. They put
on firmer ground the original suggestion by \citet{hine79} that the
edge-brightened type II \citet{fanaroff74} sources (FR~IIs) can be
distinguished into two sub-classes. They proposed a separation into
high-excitation galaxies (HEGs, defined as galaxies with
[O~III]$/$\Ha$>0.2$ and equivalent width (EW) of [O~III] $>$ 3 \AA)
and low-excitation galaxies (LEGs). \citet{tadhunter98} found a
similar result from an optical spectroscopic study of the 2Jy sample,
in which a sub-class of weak-line radio galaxies (sources with EW of
[O~III])$<10$\AA) stands out due to a low ratio between emission line
and radio luminosities as well as in the [O~III]/[O~II] line ratio.

However, despite the amount of spectroscopic data collected over the
last decades for RGs, there are still several key questions awaiting
clear answers; in particular, we must consider whether there are
indeed two (or more) distinct populations of FR~II RGs, the basis on
which they can be separated, and the relevant physical parameters
driving these distinctions.

\citet{buttiglione09,buttiglione10,buttiglione11} presented the
results of an optical spectroscopic survey of 115 3C RGs with $z<0.3$
based on observations with the Telescopio Nazionale Galileo (TNG) and
found a bimodal distribution of the relative intensity of low and high-excitation lines. Generally, HEGs are all associated with powerful radio sources
($L_{178} \gtrsim 10^{33} \ergsHz$) of FR~II morphology, while LEGs
are found at all radio power and are of both FR classes. Such a
dichotomy also corresponds to a separation in nuclear properties at
different wavelengths, with the HEGs being brighter than the LEGs (e.g.,
\citealt{chiaberge02,hardcastle06,baldi10}). This is likely related to
a different structure of the accretion disk, namely, an advection-dominated
accretion flow (ADAF, \citealt{narayan95}) in LEGs and a thin
radiatively efficient disk \citep{shakura73} (SS disk) in HEGs; as well as
to the mode of accretion (cold versus hot;
\citealt{hardcastle07}). Finally, the LEGs with the highest radio
power have FR~II morphologies indistinguishable from those seen in
HEGs \citep{baldi13}. This opens the possibility that the various
classes of RGs might be linked by an evolutionary sequence (see, e.g.,
\citealt{macconi20}). This idea is supported by the results of
\citet{balmaverde21}: the sizes and luminosities of the extended emission
line regions (EELRs) are similar in both HEGs and LEGs of the FR~II class,
indicating that they live in a similar gas rich environment. However,
there are some objections with regard to this interpretation: LEGs appear to live
in a richer environment \citep{hardcastle07,ineson15} and  less
frequently show signs of mergers \citep{ramos11,ramos12} than
HEGs. However, these differences are not conclusive enough to discard the
evolutionary scenario because there is a substantial overlap in the
properties of both the host and environment of HEGs and
LEGs. Furthermore, these might be linked to the accretion and
evolution of the radio source, thereby increasing, for instance, the fraction of LEGs
in large galaxies and in a denser environment \citep{tadhunter16}.

We have been carrying out the MUse RAdio Loud Emission lines Snapshot
(MURALES) survey of observations of 3C radio sources, with the
integral field spectrograph MUSE at the Very Large Telescope (VLT). We
initially observed 37 galaxies with $z<0.3$ and $\delta <20^\circ$
\citep{balmaverde19,balmaverde21,balmaverde22} and we have now extended MURALES to 
RGs at higher redshift observing 26 3C radio sources with $0.3 < z <
0.82$ (there are 29 such sources in the 3C with $\delta <20^\circ$,
but the observations of three of them were not obtained due to
scheduling constraints). Here, we explore the nuclear properties of
these AGNs with the main aim of deriving a spectroscopic
classification, information that is key to understanding their nuclear
properties and to unveil connections with, for instance, their radio power and
morphology. This sample extends the coverage toward higher radio power
by a factor of 10 with respect to that studied by \citet{buttiglione10}.
From the point of view of their radio structure, all 26 sources have a
FR~II radio morphology with a size exceeding $\sim$ 30 kpc, with the
only exception of 3C~138, a compact steep spectrum source with a
triple structure extending for only $\sim$6 kpc \citep{akujor93}.

These new observations also allow us to further test the unified model
(UM) for radio-loud AGNs. The UM postulates that different classes of
objects can be unified in a single population, differing solely
  for their orientation with respect to our line of sight (see, e.g.,
  \citealt{antonucci93} for a review). The origin of the aspect
dependent classification is due to the presence of: i) circumnuclear
absorbing material that produces selective absorption when the source
is observed at a large angle from its radio axis; ii) Doppler boosting
associated with relativistic motions in AGN jets. Among the several
pieces of evidence in favor of the UM, the most convincing one is the
detection of broad lines in the polarized spectra of narrow-line
objects \citep{antonucci82,antonucci84} interpreted as the result of
scattered light from an otherwise obscured nucleus. The UM has
  very successfully explained the observed properties of, for instance, BL Lac
  object and type I \citet{fanaroff74} sources, as well as of narrow and
  broad lined FR~IIs (\citealt{urry95} for a review). \citet{lawrence91},
\citet{hill96} and \citet{willott00} found that the fraction of
narrow-line objects decreases with radio power. They proposed that
this is due to a lower covering factor of the circumnuclear absorption
structure in the more luminous sources, the so-called ``receding
torus'' model.

\citet{laing94} have pointed out that LEGs should be excluded from a
sample while testing the unified model as they represent a separate
population from the HEGs (see also \citealt{wall97} and
\citealt{jackson99}). Therefore, in order to test the validity of the
UM for RL AGNs, it is necessary to separately treat  FR~II HEGs
(including those showing broad permitted lines, i.e., the broad line
objects, BLOs) and LEGs. The 3C catalog
\citep{bennett62a,bennett62b,spinrad85} is perfectly suited for probing
the validity the UM because it is based on the low-frequency radio
emission (namely, 178 MHz), free from orientation
biases. \citet{baldi13} already tested the UM on the 3C sources with
$z<0.3$ and we are now in the position of extending this analysis to
larger redshift and radio power.

However, at redshift $z \gtrsim 0.4$, not all diagnostic lines (in
particular \Ha\ and [N~II]) are covered by optical spectra and
alternative diagnostics must be used. \citet{lamareille10} considered
the spectra of a sample of emission-line galaxies of intermediate
redshift and classified them in sub-classes, star forming (SF),
LINERs, and Seyferts, based on the standard DDs. These sources were
then used to define the boundaries between these classes in a diagram
comparing the [O~III]$\lambda5007$/\Hb\ and
[O~II]$\lambda3727$/\Hb\ ratios. This method appears to be quite
robust, leading to a classification generally consistent with that
derived from the standard DDs for a large fraction of galaxies,
paving the way for the spectral classification of sources at a higher
redshift.

\begin{table}
\caption{Observations log.}
\begin{tabular}{l l r r r }
\hline
     Name      &  z    &  Obs. date & Exp. time  & Seeing  \\
\hline                                                    
     3C~016    & 0.406 &    Nov 05 2020  & 1794 &       0.49     \\ 
     3C~093    & 0.358 &    Nov 09 2020  & 1274 &       0.82     \\ 
     3C~099    & 0.426 &    Dec 08 2020  & 1674 &       0.44     \\ 
     3C~107    & 0.785 &    Dec 08 2020  & 2848 &       0.40     \\ 
     3C~109    & 0.305 &    Oct 27 2020  & 1274 &       0.38     \\ 
     3C~114    & 0.815 &    Dec 09 2020  & 2848 &       0.95     \\  
     3C~138    & 0.759 &    Dec 08 2020  & 2$\times$1780 &      0.48     \\  
     3C~142.1  & 0.406 &    Dec 17 2020  & 1794 &       0.55     \\  
     3C~175    & 0.768 &    Dec 21 2020  & 2$\times$1780 &      0.45     \\  
     3C~187    & 0.465 &    Dec 10 2020  & 1394 &       0.58     \\   
     3C~207    & 0.684 &    Dec 20 2020  & 2848 &       0.78     \\  
     3C~215    & 0.411 &    Nov 14 2020  & 1794 &       0.37     \\  
     3C~225B   & 0.582 &    Feb 18 2021  & 2248 &       0.90     \\  
     3C~226    & 0.817 &    Mar 06 2021  & 2848 &       0.83     \\  
     3C~228    & 0.552 &    Jan 20 2021  & 2248 &       0.63     \\  
     3C~275    & 0.480 &    Apr 07 2021  & 2248 &       0.50     \\
     3C~275.1  & 0.557 &    Feb 13 2021  & 2248 &       0.65     \\ 
     3C~277.2  & 0.766 &    Mar 13 2021  & 2848 &       0.57     \\  
     3C~293.1  & 0.709 &    Mar 13 2021  & 2848 &       0.53     \\
     3C~306.1  & 0.441 &    Feb 24 2021  & 1794 &       1.30     \\ 
     3C~313    & 0.461 &    Mar 12 2021  & 1794 &       0.41     \\
     3C~327.1  & 0.462 &    Mar 18 2021  & 1794 &       1.00     \\  
     3C~334    & 0.555 &    Mar 13 2021  & 2248 &       0.48     \\  
     3C~434    & 0.322 &    Nov 05 2020  & 1354 &       0.53     \\  
     3C~435    & 0.471 &    Nov 05 2020  & 1794 &       0.81     \\   
     3C~455    & 0.542 &    Nov 08 2020  & 2248 &       0.61     \\   
\hline
\end{tabular}

\medskip
Column description: (1) source name (2) redshift; (3) date of the
observation; (4) exposure time [s]; (5) mean seeing of the
observation (in arcseconds).
\label{tab1} 
\end{table}

\begin{table*}
\caption{Multiwavelength properties of the 3C sources with $0.3<z<0.82$.}
\begin{tabular}{l | l c c c | c c c c | c }
\hline 
Name      & z         &L$_{178}$&L$_{\rm [O~III]}$ & A$_{\rm V}$ & H$\beta$  & [O II] & [Ne III] & [Ne V] & Class \\
\hline
3C~016.0 & 0.405 & 34.71 & 41.06 & 0.182 & 0.567 ( 2) & 2.332 ( 4) &  0.230 ( 5) & 0.063 (29) & LEG \\
3C~093.0 & 0.357 & 34.70 & 42.20 & 0.668 & 0.104 ( 1) & 0.143 ( 5) &  0.106 ( 6) & ---        & BLO \\
3C~099.0 & 0.426 & 34.74 & 43.36 & 1.057 & 0.114 ( 1) & 0.280 ( 1) &  0.089 ( 1) & 0.033 ( 1) & HEG \\
3C~107.0 & 0.784 & 35.31 & 41.45 & 0.442 & 0.456 ( 2) & 2.850 ( 3) &  0.140 ( 2) & 0.038 (42) & LEG \\
3C~109.0 & 0.306 & 34.73 & 43.36 & 1.608 & 0.081 ( 1) & 0.084 ( 1) &  0.076 ( 1) & ---        & BLO \\
3C~114.0 & 0.815 & 35.13 & 42.73 & 1.543 & 0.098 ( 1) & 0.179 ( 4) &  0.084 ( 4) & 0.111 ( 1) & HEG \\
3C~138.0 & 0.755 & 35.59 & 43.50 & 0.799 & 0.145 ( 1) & 0.105 ( 1) &  0.090 ( 1) & 0.031 ( 1) & BLO \\
3C~142.1 & 0.406 & 34.95 & 41.86 & 0.916 & 0.114 ( 2) & 0.722 ( 2) &  0.103 ( 4) & 0.047 ( 4) & HEG \\
3C~175.0 & 0.770 & 35.51 & 43.40 & 0.398 & 0.089 ( 1) & 0.149 ( 1) &  0.138 ( 1) & 0.055 ( 1) & BLO \\
3C~187.0 & 0.467 & 34.70 & 41.30 & 0.210 & 0.080 ( 4) & 0.719 ( 2) &  0.159 ( 6) & 0.084 (16) & HEG \\
3C~207.0 & 0.680 & 35.28 & 42.80 & 0.265 & 0.079 ( 1) & 0.116 ( 1) &  0.114 ( 1) & 0.105 ( 1) & BLO \\
3C~215.0 & 0.411 & 34.73 & 42.68 & 0.109 & 0.103 ( 1) & 0.123 ( 1) &  0.122 ( 1) & 0.140 ( 1) & BLO \\
3C~225.0 & 0.583 & 35.33 & 42.35 & 0.151 & 0.255 ( 1) & 1.415 ( 1) &  0.147 ( 2) & 0.031 ( 5) & LEG \\
3C~226.0 & 0.817 & 35.50 & 42.79 & 0.074 & 0.112 ( 1) & 0.254 ( 1) &  0.074 ( 1) & 0.074 ( 1) & HEG \\
3C~228.0 & 0.552 & 35.29 & 42.08 & 0.092 & 0.139 ( 1) & 0.660 ( 1) &  0.113 ( 2) & 0.068 ( 3) & HEG \\
3C~275.0 & 0.478 & 34.98 & 42.34 & 0.067 & 0.103 ( 1) & 0.213 ( 2) &  0.140 ( 2) & 0.196 ( 1) & HEG \\
3C~275.1 & 0.556 & 35.22 & 42.88 & 0.076 & 0.086 ( 1) & 0.397 ( 1) &  0.089 ( 2) & 0.037 ( 1) & BLO \\
3C~277.2 & 0.765 & 35.34 & 43.09 & 0.067 & 0.077 ( 1) & 0.126 ( 7) &  0.065 ( 9) & 0.059 ( 1) & HEG \\
3C~293.1 & 0.709 & 35.15 & 41.00 & 0.058 & 0.259 ( 3) & 1.791 ( 4) &  0.147 (28) & 0.138 (24) & LEG \\
3C~306.1 & 0.441 & 34.87 & 42.59 & 0.274 & 0.091 ( 1) & 0.131 ( 2) &  0.063 ( 2) & 0.079 ( 1) & HEG \\
3C~313.0 & 0.459 & 35.09 & 41.82 & 0.084 & 0.114 ( 1) & 0.645 ( 5) &  0.096 ( 9) & 0.111 ( 3) & HEG \\
3C~327.1 & 0.462 & 35.16 & 42.39 & 0.364 & 0.107 ( 1) & 0.280 ( 1) &  0.144 ( 1) & 0.081 ( 1) & HEG \\
3C~334.0 & 0.555 & 34.99 & 43.29 & 0.113 & 0.042 ( 1) & 0.197 ( 1) &  0.156 ( 1) & 0.043 ( 2) & BLO \\
3C~434.0 & 0.323 & 34.13 & 41.14 & 0.255 & 0.332 ( 3) & 1.496 ( 5) &  0.210 ( 4) & ---        & LEG \\
3C~435.0 & 0.471 & 34.87 & 41.46 & 0.163 & 0.253 ( 1) & 2.006 (12) &  0.177 (30) & 0.081 (10) & LEG \\
3C~455.0 & 0.543 & 35.04 & 42.76 & 0.138 & 0.119 ( 1) & 0.448 ( 4) &  0.140 ( 3) & 0.041 ( 2) & HEG \\
\hline
\end{tabular}

\medskip
Column descriptions: (1) name; (2) redshift, (3) logarithm of the
radio luminosity at 178 MHz [erg s$^{-1}$ Hz$^{-1}$] derived from the
fluxes listed by \citet{spinrad85} and K-corrected assuming a radio
spectral index of 0.7; (4) logarithm of the [O~III]$\lambda$5007
luminosity [erg s$^{-1}$], corrected for the Galactic absorption; (5)
Galactic A$_{\rm V}$; (6 through 9) de-reddened flux ratios of the
diagnostic lines with respect to [O~III]. The values in parentheses
report the errors (in percentage) of each line; (10) spectroscopic
classification, as derived in Sect. \ref{bDD}.
\label{tab2} 
\end{table*}

Here, we follow a similar approach with the aim of exploring the nuclear properties of
the 3C sources. With respect to the analysis of \citet{lamareille10},
we consider  additional forbidden lines, namely
[Ne~III]$\lambda3870$ and [Ne~V]$\lambda3426$, testing whether they
can be used to obtain a robust spectroscopic classification of these
sources into the different classes. These emission lines are closer in
wavelength to the [O~II]$\lambda\lambda3727$ doublet and this reduces
the uncertainties related to the reddening correction.

The paper is organized as follows. In Sect. \ref{observation}, we
present the observations of the 26 RGs at intermediate redshift
(izRGs) and describe the data reduction. In Sect. \ref{bDD}, we explore
the ability of the blue diagnostic diagrams, based only on lines at
wavelengths shorter than the [O~III]$\lambda5007$ line, to separate
the various sub-classes of emission line galaxies and we derive a
classification for the sources in the present sample. In
Sects. \ref{ropt} and \ref{UM}, we study the connection between radio
power and line emission and consider the implications of our findings
for the unified model of radio-loud AGNs. In Sect. \ref{discussion}, we
discuss our results and in Sect. \ref{summary}, we provide a summary
and our conclusions.

We adopt the following set of cosmological parameters: $H_{\rm
  o}=69.7$ \kms\ Mpc$^{-1}$ and $\Omega_m$=0.286
\citep{bennett14}. The results from the literature used for comparison
are all scaled to this cosmology.

\section{Observations and data reduction}
\label{observation}

\begin{figure*}
  \includegraphics[width=0.49\textwidth]{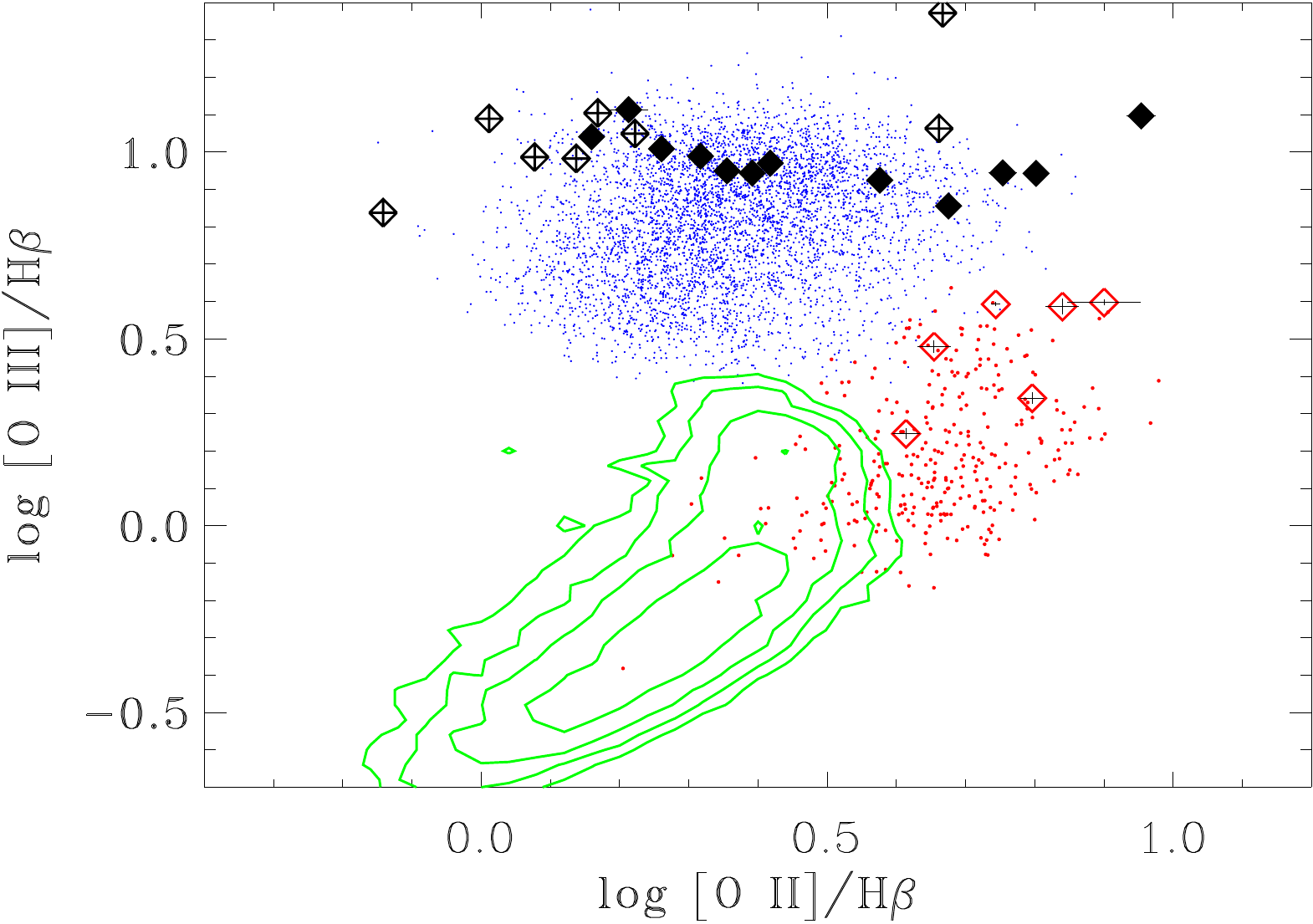}
  \includegraphics[width=0.49\textwidth]{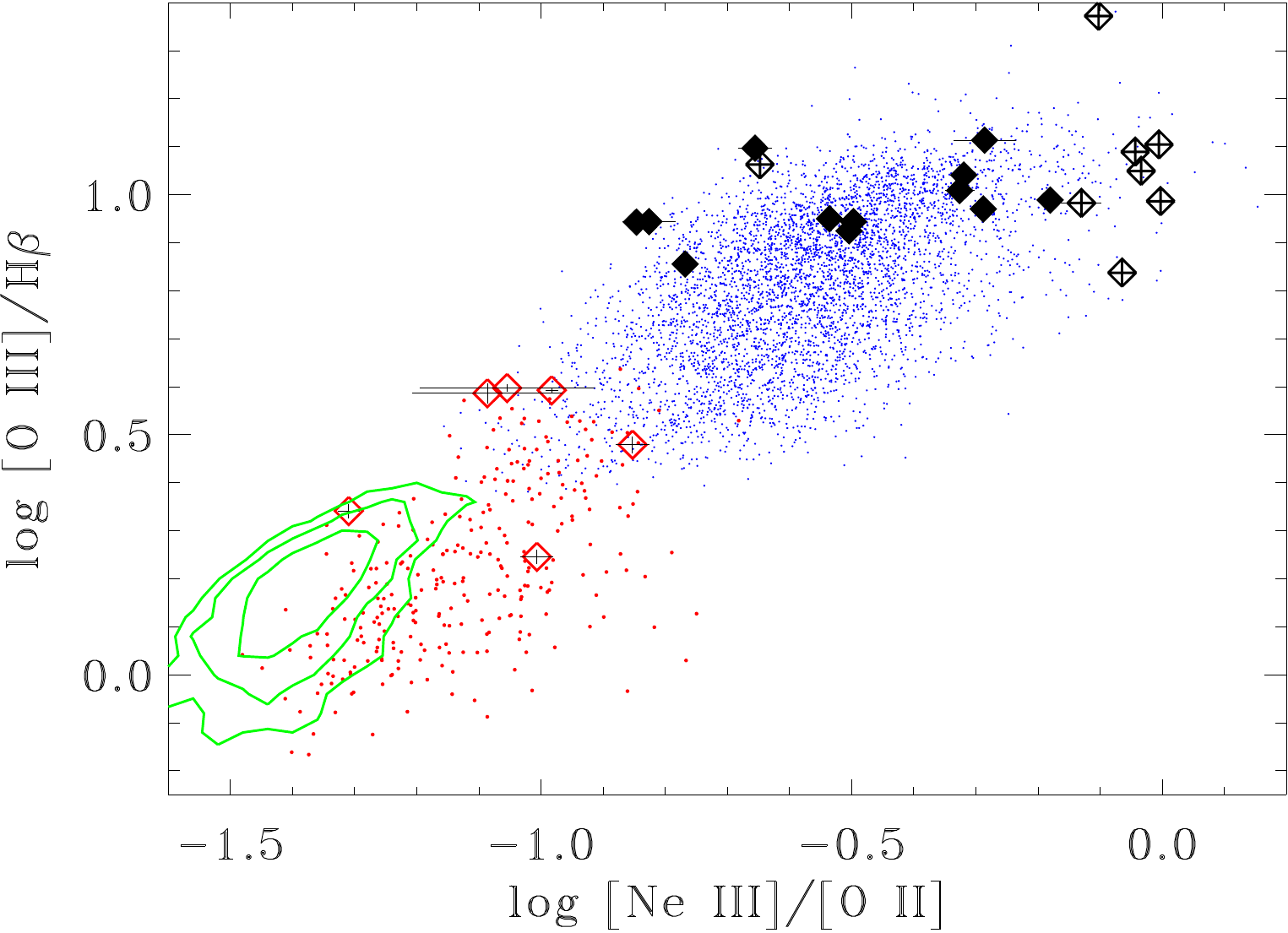}
  \includegraphics[width=0.49\textwidth]{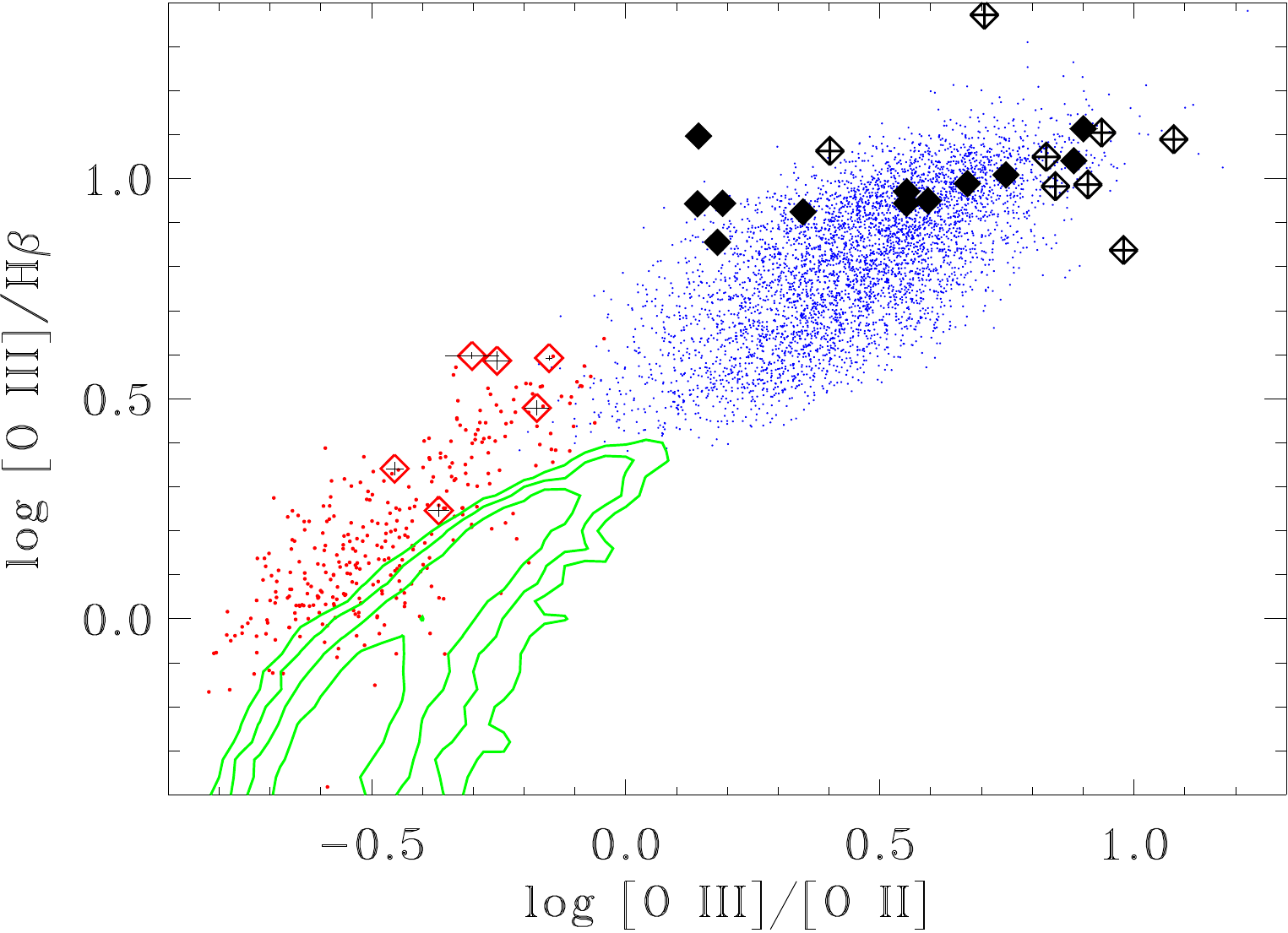}
  \includegraphics[width=0.49\textwidth]{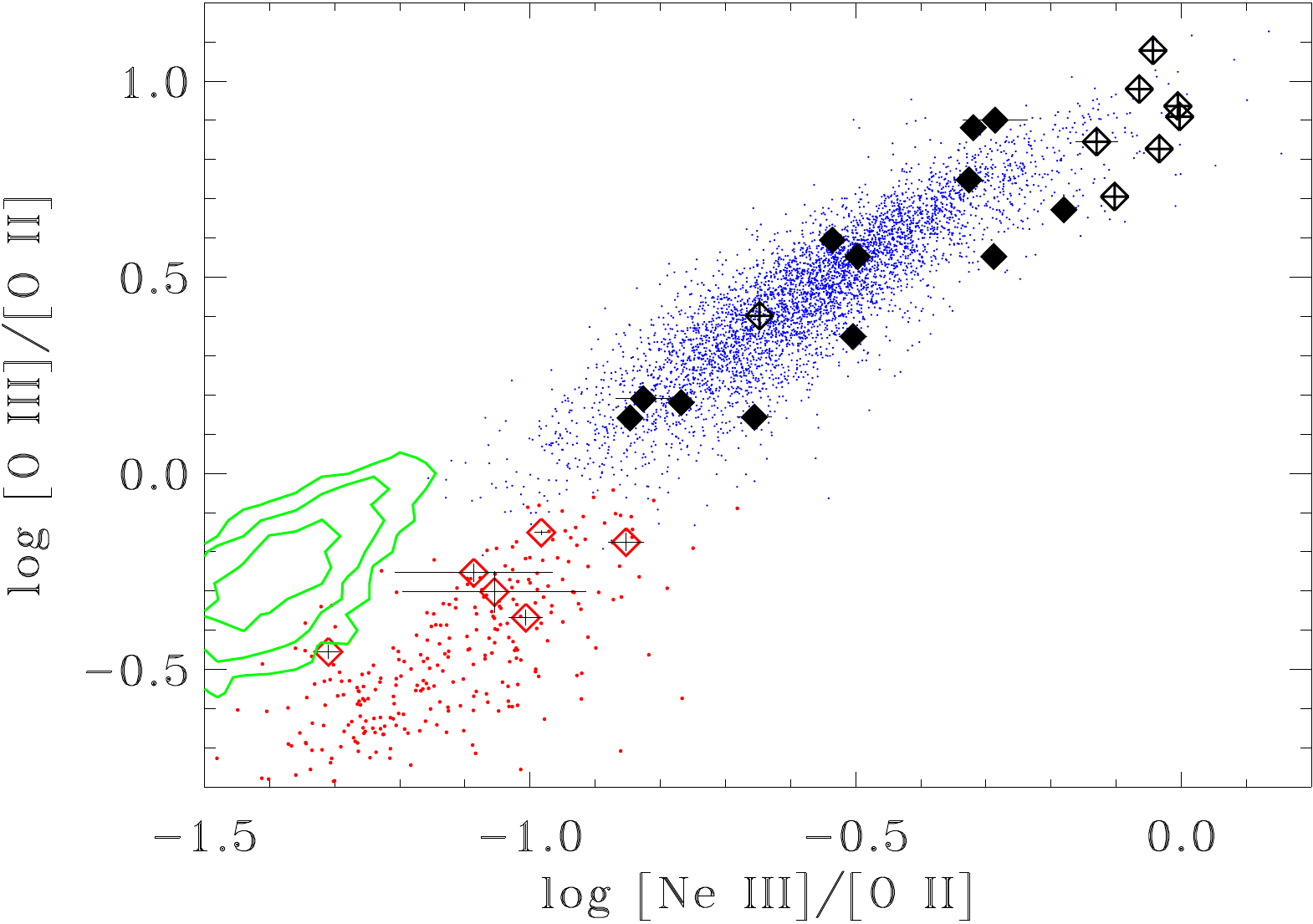}
\caption{Diagnostic diagrams obtained considering only emission lines
  with rest frame wavelength $\lambda \leq 5007$\AA\ for the emission
  line galaxies in the SDSS/DR7. The blue and red dots represent the
  location of Seyferts/HEGs and LINERs/LEGs, respectively, selected
  based on their location in the [O~III]/\Hb\ versus
  [S~II]/\Ha\ diagram. The green contours are the iso-densities of the
  location of SF galaxies. The diamonds are the 26 3C at 0.3 $<$
    z $<$ 0.82 observed with MUSE. The black filled diamonds are the
    izRGs classified as HEGs, the diamonds surrounding a cross are the
    BLOs, while the red ones are the LEGs.}
\label{blue2bis}
\end{figure*}

The 26 izRGs were observed between October 2020 and April 2021 with
the MUSE spectrograph. The observing times increase with redshift and
range from 1274 to 2848 s, see Tab. \ref{tab1}.\footnote{3C~138 and
  3C~175 were observed twice with an exposure time of 1780 s.} We used
the ESO MUSE pipeline (version 1.6.2) to obtain a fully reduced and
calibrated data cube. We extracted the spectrum from a synthetic
aperture of 2\arcsec $\times$ 2 \arcsec. This is the same extraction
aperture used by \citet{buttiglione09} but in the izRGs covers a
region with size between 9 and 15 kpc. We expect that the larger
  physical size of the extraction aperture has a limited effect, due
  to the dominance of the small scale emission line fluxes in all
  classes of RGs and, as we show later, to the similarity of the line
  ratios for the central and extended regions. The resulting spectra
are shown in Appendix \ref{spectra}. In several of these, stellar
absorption lines are clearly visible. In these cases we subtracted the
stellar continuum by using stellar population models based on the
MILES stellar templates library \citep{vazdekis10} and the Penalized
Pixel-Fitting code (pPXF, \citealt{cappellari04,cappellari17}). In the
remaining sources, we modeled the featureless continuum with a power
law.

We then fitted the brightest emission-lines, namely
[Ne~V]$\lambda3426$, [O~II]$\lambda\lambda3726,3729$,
[Ne~III]$\lambda3870$, H$\beta$, [O~III]$\lambda\lambda$4959,5007, in
the continuum subtracted spectra. For sources having low signal-to-noise ratio (S/N)
lines, we assumed that those close in wavelength (i.e., the two groups
in the UV and blue portion of the spectra) have the same velocity
profile. This is actually only the case for the [Ne V] line. In the
sources with the lowest redshift, also the
[O~I]$\lambda\lambda$6300,6363, H$\alpha$, and
[N~II]$\lambda\lambda$6548,6584 lines are accessible and their
intensities were also measured. The relative intensities of the lines
of the [O~III] and [N~II] doublets were fixed to their atomic ratio of
three. A single Gaussian rarely reproduces accurately the lines
profiles and in most galaxies we included a second Gaussian component,
with the same constraints described above. In the eight sources marked
as broad line objects (BLOs), in Table \ref{tab2} we include an
additional broad component for the \Ha\ (when visible) and
\Hb\ lines. In three sources (namely, 3C~138, 3C~175, and 3C~334),
there is a significant contribution from iron lines. This component is
removed from the spectra by using the Fe templates (properly scaled
and broadened) of \citet{vestergaard01}.

The line ratios against the [O~III] line are tabulated in
Tab. \ref{tab2}. These values are corrected for Galactic absorption
adopting the parameterization from \citet{fitzpatrick99}. For the
sources where a broad component in the Balmer lines is present, the
ratios only refer  to the narrow components.

\section{Exploring new diagnostic diagrams}
\label{bDD}

\subsection{Blue diagnostic diagrams}

\citet{lamareille10} showed that the [O~III]/\Hb\ versus
      [O~II]/\Hb\ diagram is a useful tool for classifying emission line
      galaxies at moderate redshift with optical spectra (thus not
      covering the [N~II] and \Ha\ spectral region) into star-forming,
      Seyferts, and LINERs. We here follow a similar approach and
      extend it to include additional emission line ratios. We define
      these diagnostic diagrams as blue-DDs.

In Fig. \ref{blue2bis} (top-left panel), we show the location in the
blue-DDs of the emission line galaxies included in the main sample of
$\sim$800000 sources with spectra available from the SDSS
\citep{york00}, Data Release 7, by using the MPA-JHU release of
spectral measurements\footnote{Available at {\sl
    http://www.mpa-garching.mpg.de/SDSS/DR7/}}. We selected a
sub-sample of sources in which all lines of interest are covered by
the SDSS spectrum (i.e., with $0.03 < z < 0.35$) and are detected with
a significance of at least 5$\sigma$. We separated them into the three
classes (Seyferts, LINERs, and SF) based on their location in the
[O~III]/\Hb\ versus [S~II]/\Ha\ diagram where they are best
separated. We left aside the sources located at less than 0.1 dex from
the boundaries between the various classes, to exclude objects of
ambiguous classification. The three groups populate different regions
of [O~III]/\Hb\ versus [O~II]/\Hb\ diagram with only marginal
overlaps. As shown by \citet{curti17}, the SF sequence corresponds to
an increasing metallicity moving towards lower
[O~III]/\Hb\ ratios. Out of the 26 3C izRGs (the black symbols in this
figure), six are located among the LINERs/LEGs, while the eight BLOs
are located into the region of Seyferts/HEGs; the remaining sources
are located among the Seyferts/HEGs.

This diagram has various disadvantages with respect to the traditional
DDs: 1) the quantities on both axis depend on the \Hb\ line, a
relatively weak line whose measurement is often difficult due to the
presence of stellar absorption (in faint AGNs) or of a broad component
(in type I AGNs); 2) the relatively large separation between the
[O~II] and \Hb\ lines causes their ratio to depend on the correction
for dust absorption. We then experimented with alternative blue-DDs in
which one or both of these issues are overcome. Three of these
diagrams are presented in Fig. \ref{blue2bis}. Overall, they confirm
the results obtained from the comparison of the [O~III]/\Hb\ and
[O~II]/\Hb\ ratios. In particular in the two bottom panels (showing
[O~III]/\Hb\ versus [O~III]/[O~II] and [O~III]/[O~II] versus
[Ne~III]/[O~II], respectively) the different spectroscopic classes
among the SDSS galaxies are best separated. The 26 izRGs form two
distinct groups and their location into the different regions of the
diagrams is consistently maintained.

These diagrams thus enable us to classify them into the different
spectroscopic classes, as reported in Tab. \ref{tab2}: six are LEGs
and 20 are HEGs. Eight of the HEGs show the presence of broad
\Hb\ lines: we then classify them as BLOs. This confirms the
conclusion of \citet{buttiglione10} that the narrow lines in BLOs have
line ratios similar to those of high-excitation galaxies.

In the six izRGs at the lowest redshift ($z \lesssim 0.4$), the MUSE
spectra cover the [N~II]+\Ha\ complex (see Fig. \ref{redspectra}). For
these objects, it is possible to derive a classification based on the
traditional DDs as a sanity check for the results based on the
blue-DDs. These are 3C~016 (LEG), 3C~093 (BLO), 3C~109 (BLO), 3C~142.1
(HEG), 3C~215 (BLO), and 3C~434 (LEG). Their line ratios are listed in
Table \ref{tabred}. As shown by Fig. \ref{redDD}, the classification
based on the blue-DDs is confirmed for these sources.

\begin{table}
\caption{Logarithm of the emission line ratios in the red portion of
  the MUSE spectra.}
\begin{tabular}{l c c c c c c c c c }
\hline 
Name        &[O~III]/H$\beta$ & [N~II]/\Ha & [S~II]/\Ha & [O~I]/\Ha \\
\hline
3C~016.0 & 0.24 (2) & -0.23  (1) & ---        & -0.49  (1) \\
3C~093.0 & 0.98 (1) & -0.11  (2) & -0.77  (2) & -0.84  (3) \\
3C~109.0 & 1.09 (1) & -0.99  (1) & -0.92  (3) & -1.33  (1) \\
3C~142.1 & 0.94 (2) & -0.08  (1) & ---        & -0.55  (1) \\
3C~215.0 & 0.98 (1) & -0.24  (1) & ---        & -0.61  (1) \\
3C~434.0 & 0.48 (3) & -0.04  (1) & -0.34  (3) & -0.52  (1) \\
\hline
\end{tabular}

\medskip
Columns description: (1) name; (2-4) logarithm of the diagnostic line
ratios. In Fig. \ref{redDD}, we show the standard DDs for these sources. 
\label{tabred} 
\end{table}

\subsection{Diagnostic diagrams from UV lines}
\label{uvDD}

\begin{figure}
  \includegraphics[width=0.49\textwidth]{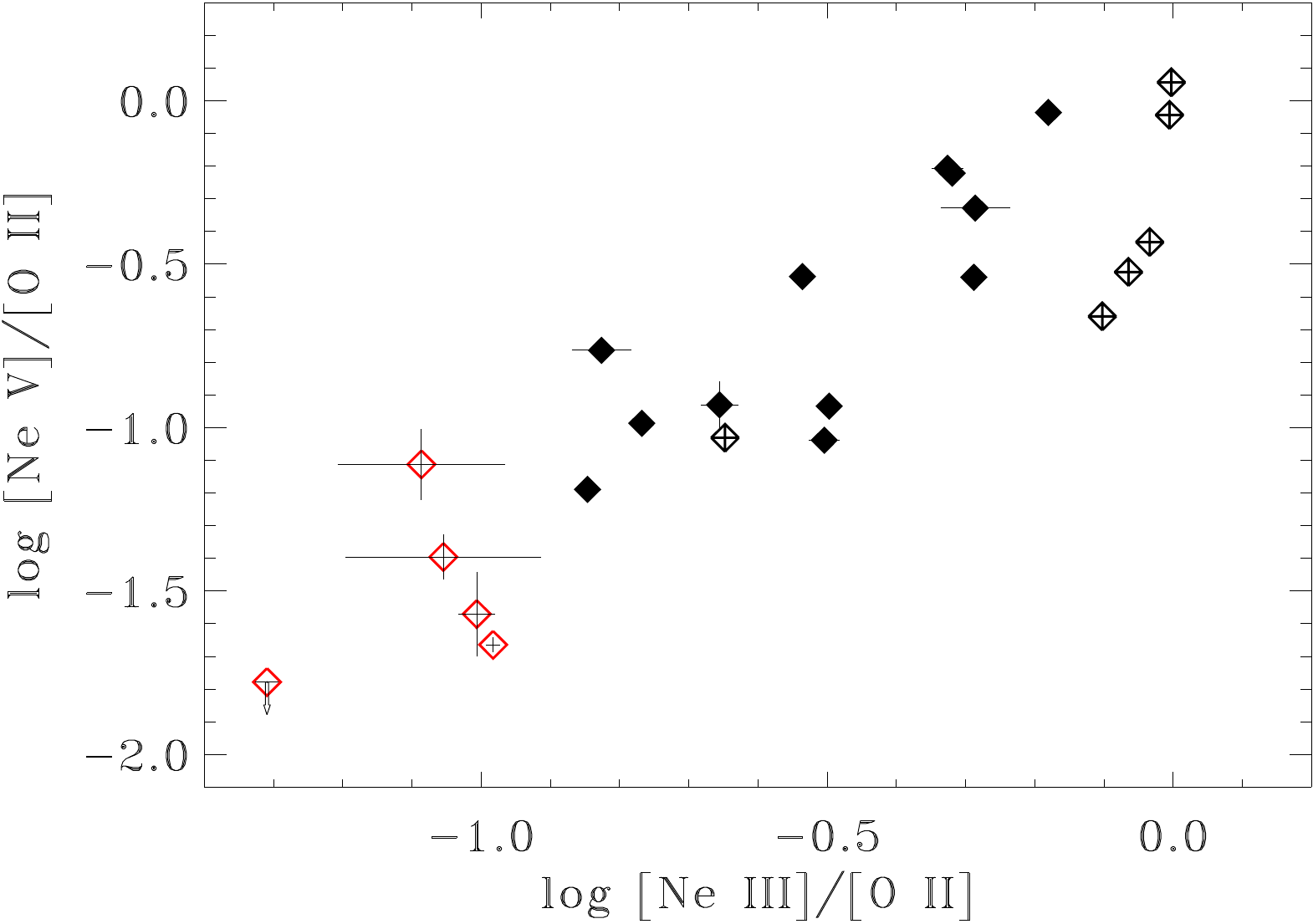}
\caption{Diagnostic diagram obtained from line ratios, considering
  only emission lines with a rest frame wavelength of $\lambda \leq
  3870$\AA\ for the 23 RGs with $z>0.4$ (i.e., those for which the
  MUSE spectra include the [Ne~V] line). The black filled diamonds are
  the HEGs, the diamonds surrounding a cross are the BLOs, and the red
  ones are the LEGs.}
\label{uv}
\end{figure}

Having obtained a spectroscopic classification based on the blue-DDs,
we can explore new diagrams based only on UV lines, namely,
[Ne~V]$\lambda$3426, [O~II]$\lambda$3727, and [Ne~III]$\lambda3870$,
the UV-DDs. In Fig. \ref{uv} we report the location of the izRGs into
the [Ne~V]/[O~II] versus [Ne~III]/[O~II] plane. This is possible for
all but the three sources at the lowest redshift whose MUSE spectra do
not include the [Ne~V] line. The sources classified as LEGs from the
blue-DDs are all located in the bottom left corner of this diagram,
that is, they have a relatively brighter low-ionization line with respect
to the HEGs and BLOs, and they are adequately separated from the objects belonging to
these two spectroscopic classes. We conclude that the UV-DDs
provide us with a further robust tool for obtaining a spectroscopic AGN
classification.

The importance of this UV-DDs is two-fold: it is based only on lines
that are very close in wavelength, making it robust against uncertainties in
the correction for dust absorption, which is particularly important in this
spectral region. Secondly, these three lines are visible in optical
spectra out to a redshift $z\sim1.5$ and this makes this diagram a
unique tool for the spectral classification of distant sources.

\section{Radio and emission lines connection}
\label{ropt}

\citet{buttiglione10} found, based on the subsample of 3C sources with
z$<$0.3, a clear trend of increasing line luminosity with radio
power. This radio-optical connection was already found and discussed
by several authors (e.g.,
\citealt{baum89a,baum89b,rawlings89,rawlings91,willott99}). By
separating the different spectroscopic classes,
\citeauthor{buttiglione10} noted that while LEGs cover the whole range
of radio power, HEGs are only found at radio luminosities larger than
$L_{178} \gtrsim 10^{33} \ergsHz$. By considering separately the
sub-populations of HEGs and LEGs, they found that both classes obey to
a quasi-linear correlation between $L_{\rm[O~III]}$ and $L_{178}$,
with the HEGs having a line luminosity a factor of $\sim 10$ higher than
LEGs in the common range of radio power.

\begin{figure}
  \includegraphics[width=0.49\textwidth]{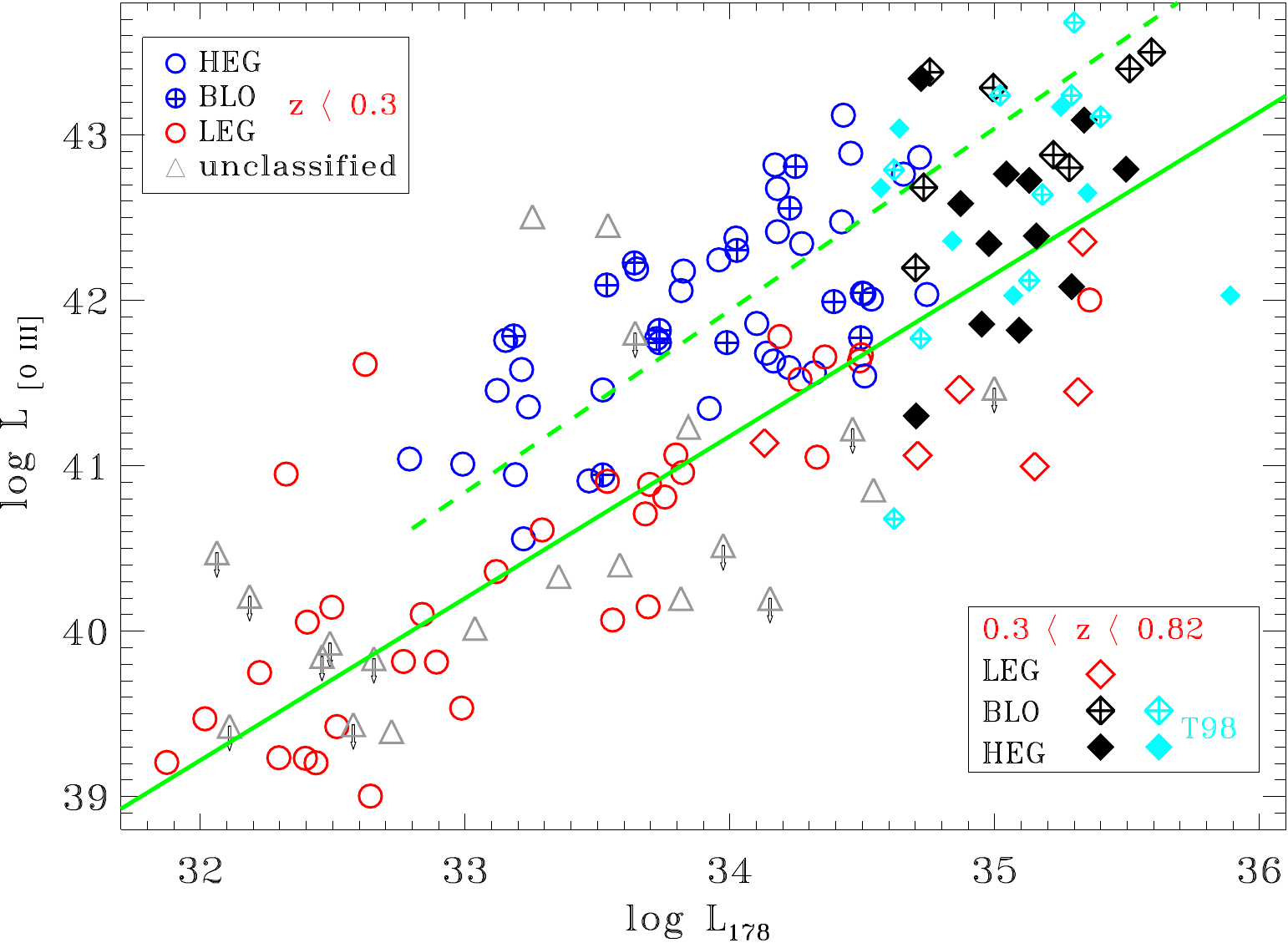}
\caption{[O~III] (in erg s$^{-1}$ units) emission line versus radio
  luminosity (in $\ergsHz$ units) for the 3C sources, separated into
  the various spectroscopic classes. The data for the objects at
  z$<$0.3 are from \citet{buttiglione10,buttiglione11}, those at
  higher redshift are from our analysis. The two green lines represent
  the linear correlation between radio power and line luminosity
  derived by \citeauthor{buttiglione10} for HEGs (dashed) and LEGs
  (solid) separately. The cyan symbols represent the izRGs from
    \citet{tadhunter98} added to our sample to improve the statistical
    significance of our analysis.}
\label{l178lo3}
\end{figure}

In Fig. \ref{l178lo3}, we compare the [O~III] line luminosity with the total
radio luminosity at 178 MHz of the 3C sources from
\citet{buttiglione10} and including the new measurements for the 26 3C
izRGs. As expected, because the 3C is a flux limited sample, the
higher redshift sources are in general those of higher radio power,
extending the coverage to $L_{178} \sim 3 \times 10^{35} \ergsHz$. We
confirm the finding that HEGs and BLOs have larger line luminosities
than LEGs at similar radio power. It is of great interest the result
that LEGs are also found  among the sources of highest radio power and we
 return to this topic  in Sect. \ref{discussion}.

Overall, the eight BLOs and the 12 HEGs are located at higher line
luminosities than the lower redshift 3C. However, the majority of
these sources (17 out of 20) lie below the extrapolation of the
$L_{\rm[O~III]}$ - $L_{178}$ relation of the objects of these same
spectroscopic classes at low-z. Considering their median radio (log
$L_r$[erg/s]$ = 35.07$), the expected line luminosity is a factor of 2.5
higher than the observed median value (log $L_{\rm[O~III]} $[erg/s]$=
42.74$). A similar effect is found when considering the LEGs. Although
the trend of increasing line luminosity with radio power is still
present, the slope of radio-line decreases at the highest power.

To improve the statistical significance of this result we included in
the analysis the spectroscopic study presented by
\citet{tadhunter98}. In their sample there are 16 izRGs, nine BLOs and
seven HEGs based on their [O~III]/[O~II] ratios. Eleven izRGs of this
sample are located below the extrapolation of the $L_{\rm[O~III]}$ -
$L_{178}$ relation, strengthening the statistical significance of its
decreased slope at high luminosities.

\citet{croom02} found a similar effect in quasi-stellar objects (QSOs)
where a decrease in the equivalent width of the narrow lines with
increasing nuclear luminosity is observed. This is reminiscent of the
Baldwin effect that is, however, seen in the permitted UV emission
lines \citep{baldwin77,baldwin89}. \citeauthor{croom02} argued that
this behavior is due to the NLR dimensions becoming larger than the
host galaxy. This effect is directly seen from NLR imaging of type II
quasars that show a flattening of the relationship between NLR size
and AGN luminosity \citep{hainline13,hainline14}. These results do not
imply that there is a lack of gas at larger distances (indeed,
ionized gas is seen in most of the RGs studied on scales of several
tens of kiloparsec) but they indicate that in more luminous AGNs a
higher fraction of nuclear ionizing photons escape without being
absorbed and reprocessed by the interstellar medium (ISM).

\section{Unified model for radio-loud AGNs.}
\label{UM}

As discussed in the Introduction, the unified model for radio-loud
AGNs is based on the idea that the different sub-classes, for instance,
  broad- and narrow-lined FR~II, differ only with respect to their orientation
with respect to our line of sight. Here, we limit our analysis to the
optical spectroscopic classification: BLOs and HEGs share
similar narrow-line properties and differ mainly for the presence
of an obscuring structure hiding the broad line region (BLR) in the
HEG class. LEGs must be instead considered as a intrinsically
different sub-class as already pointed out by \citet{wall97} and
\citet{jackson99}.

In the sample of 3C radio sources with $z<0.3,$ \citet{baldi13} found
18 BLOs and 33 HEGs, implying that nuclear obscuration hides the BLR
in $\sim$65\% of the objects: this fraction is independent of
redshift, up to $z=0.3$, and radio power, up to $L_{178} \sim 4\times
10^{34} $ $\ergsHz$. In the present sample, we have 12 HEGs and 8
BLOs, with a fraction of the HEGs (40\%) being slightly smaller than in lower redshift
sources. Nonetheless, a test of equal proportions \citep{edwin27}
indicates that there is no evidence for a statistically significant
difference between the fraction of broad- and narrow-line sources below
and above $z=0.3$: a larger sample of izRGs is required to properly
test the receding torus model, that is, the possibility that in the more
luminous sources, the circumnuclear absorption structure has a lower
covering factor.

Looking in more detail at the narrow line properties of HEGs and BLOs,
the latter class shows generally higher [Ne~V]/[O~II],
[Ne~III]/[O~II], and [O~III]/[O~II] ratios than HEGs. A similar effect
has been noted by \citet{baldi13} by comparing the luminosity of the
[O~III] and [O~II] lines of 3C RG with z$<$0.3. They interpreted this
result as due to the presence of a narrow line emitting region with a
density higher than the [O~II] critical density located within the
walls of the obscuring torus: this region is visible in BLOs but is
obscured in HEGs, leading to a difference in the ratios of these
emission lines. The same conclusion applies to the comparison of the
relative strength of the BLOs and HEGs among the 3C izRGs because,
similarly to [O~III], also the [Ne~V] and [Ne~III] have a larger
critical density than [O~II].\footnote{The critical densities are
  1.6$\times10^7$, 4.5$\times10^3$, 9.7$\times10^6$, and
  7.0$\times10^7$ cm$^{-3}$ for [Ne~III], [O~II], [Ne~V], and [O~III],
  respectively \citep{derobertis86}.} The presence of a density
stratification within the NLR is also supported by the larger widths that are generally observed in AGN, coming from the forbidden lines with higher critical
density produced by ionized gas located at smaller distances from the
supermassive black hole (e.g., \citealt{balmaverde14}). The
  presence of a portion of the NLR hidden by the torus was suggested
  also by \citet{haas05} for radio-galaxies and \citet{baum10} for
  Seyfert galaxies. The connection between line and radio luminosity
  is also affected by the presence of this high density component of
  the NLR, visible only in BLOs: the median $L_{[O~III]}/L_{178}$
  ratio is in fact $\sim$ 30\% higher in BLOs than in HEGs.

Concerning the population of LEGs, in the izRGs sample we have 6 LEGs,
corresponding to 23\%; below $z=0.3$ we have 21 LEGs with a FR~II
morphology, 29\% of the edge-brightened sources. The fraction of LEGs
apparently does not strongly depend on redshift and the equal
proportions test does not return a statistically significant
difference between the low- and intermediate-redshift RGs.

\section{Discussion}
\label{discussion}

The various classes of radio loud AGNs have been interpreted as the
result of different accretion rates (and accretion modes) onto the
central supermassive black hole (SMBH) related to the change of the
structure of the accretion disk and of the spectral energy
distribution of the nuclear radiation field
\citep{baum95,ghisellini01,hardcastle07,buttiglione10}.  In fact, HEGs
and LEGs differ for their nuclear properties, with HEGs/BLOs having
more luminous nuclei (both in the optical, \citealt{chiaberge02}, in
the X-ray, \citealt{hardcastle00,hardcastle06}, and in the
near-infrared, \citealt{baldi10}) and higher Eddington ratios than
LEGs of similar radio power. More specifically, LEGs are thought to be
associated with a low accretion rate of hot gas forming an ADAF-like
disk, while HEGs (and BLOs) are powered by a thin radiatively
efficient disk maintained by a high accretion rate of cold
gas. \citet{nagao02} suggested that the harder nuclear ionization
field produced by ADAFs with respect to standard accretion disks, in
particular, the larger fraction of X-ray photons, causes a relatively
higher intensity of the low ionization lines. While LEGs nuclei are
found both in FR classes, HEGs are only found in FR~IIs.

A key result of our spectroscopic analysis of the izRGs is that
FR~II/LEGs are found also at the highest level of radio power, among
the most luminous radio sources in the Universe. Furthermore, their
fraction within the FR~II RGs population does not strongly depend on
radio luminosity, being 23\% for 0.3$<z<$0.82 and 29\% of the
edge-brightened sources at $z<0.3$.

\citet{baldi13} have shown that the distributions of radio luminosity,
core dominance, and source size of the LEG and HEG/BLO classes in the
FR~II 3C sources at $z<0.3$ are not statistically different. A similar
analysis for the izRGs is plagued by the smaller sample size. We
measured the source sizes and core flux densities from the Very Large
Array Sky Survey images at 3 GHz \citep{lacy20}. The median sizes (240
kpc and 310 kpc for LEGs and HEGs, respectively) and the median power
at 178 MHz for the two classes (log L$_{178}$=35.09 and log
L$_{178}$=35.02, both in $\ergsHz$ units, for LEGs and HEGs,
respectively) are both similar. Radio core measurements are
  possible for only 3/6 6 LEGs, 6/12 HEGs, and 6/8 BLOS. This hinders
  us from drawing any robust comparison of the core dominance of the
  different spectroscopic classes among the izRGs. Furthermore, the LEGs and HEGs
cannot be readily distinguished also from a more subjective
comparison, based on the presence of hot spots or on the morphology of
the radio lobes.

The similarity of the radio properties between FR~IIs LEGs and HEGs,
contrasted by the differences in the nuclear characteristics, appear
to imply that the process of jet formation is unrelated to the disk
structure and to the accretion process. This is what is expected if
the jet launching process in powerful radio-loud AGN is due to
the extraction of the rotational energy of the supermassive black hole
\citep{blandford77,penna13}. The link between radio power and line
emission (a proxy for the AGN bolometric luminosity) indicates that,
within each spectroscopic class, a connection between accretion rate
and radio power is present.

Alternatively, there might be an evolutionary link between LEGs and
HEGs and, for instance, the transition from a HEG to a LEG (or vice versa)
might occur following a drop (or an increase) in the fuel supply
within an activity cycle. \citet{capetti11} discussed the properties
of a class of radio sources, the extremely low-excitation galaxies
(ELEGs), identified as remnant objects in which the accretion has
ceased. The ELEGs are characterized by an extremely low level of gas
excitation, a high deficit of line emission with respect to RGs of
similar radio luminosity, and a relatively low luminosity of their
radio cores. \citet{capetti13} showed that, from the point of view of
the nuclear line ratios, the transition from HEG to ELEG occurs in a
time interval smaller than the NLR light travel time, namely, $\sim 10^3$
years. The timescale during which such sources display LEG-like line
ratios is even shorter: given the duration of this phase, compared to
the lifetime of radio galaxies, remnant sources cannot contribute
significantly to the population of radio galaxies and to the LEG
population in particular. Furthermore, ELEG have a radio core
dominance much lower than the remaining population of RGs, an effect
that is not observed in the LEGs.

Instead, HEGs and LEGs might be related to episodes of high and
low accretion rate, namely, that the HEG phase is recurrent.
The various structures of RGs respond to a change of their nuclear
activity on different timescales (see \citealt{tadhunter16}, for a
more detailed discussion of this issue). In particular, the transition
from HEG to LEG in the spectrum extracted from a synthetic nuclear
aperture occurs, as already discussed, in $\lesssim 10^3$
years. However, \citet{balmaverde21} showed that ionized gas
structures extend to sizes often exceeding tens of kpc in the FR~IIs
(in both HEGs and LEGs) and with a median size of 16 kpc. The EELR
responds to changes in the nuclear ionizing field on its light travel
crossing time, $\tau_{\rm EELR} \sim 5\times 10^4$
years. \citet{balmaverde22} show that the EELR in all the 15
HEGs/BLOs with $z<0.3$ has line ratios similar to those seen in their
nuclei and corresponding to a classification as high-excitation
regions. In a forthcoming paper, where we discuss the properties of
the EELR in the izRGs, we show that the same result applies to
the izRGs. Twenty-two of them (15 of which are HEGs or BLOs) have
well-resolved EELR and the emission line intensities measured in the
off-nuclear regions lead to the same spectroscopic classification of
the nucleus (although the EELR in these higher redshift sources are
often punctuated by star forming regions). The same result applies to
9 LEGs in the two samples, for a total of $N=39$ sources. Therefore,
we have no evidence for RGs caught during the transition between HEG
and LEG, or viceversa: the minimum value for this transition timescale
is then $\gtrsim \tau_{\rm EELR} \times N \sim 2\times 10^6$ years.

The extended radio structures also change if the amount of energy
carried by the relativistic jets decreases following a transition from
HEG to LEG, that is, if the jet properties are closely related to the
accretion mode. When the jets are no longer carrying relativistic
electrons (or when the injection is strongly reduced), the lobes will
significantly suffer for adiabatic expansion and cooling after a
timescale on the order of $\sim 10^7$ years (e.g.,
\citealt{english19}). However, the hot spots will be affected on a
much shorter timescale, namely, after one light crossing time to the
edges of the radio source: considering the median distance of the hot
spots in the RGs of our sample ($\sim 150$ kpc), this is on the order of $\sim
5\times 10^5$ years. The internal sound crossing time of the hot
spots, the time corresponding to a significant expansion and
consequent drop of luminosity, is instead negligible in this context.

The tension between the transition timescale derived from the
properties of the EELR and that obtained from the presence of hot
spots in the LEGs suggests that the recurrency scenario in which HEGs
and LEGs represent the manifestation of episodes of high and low
accretion rates, also related to a change in the structure of the
accretion disk, is highly contrived.

\section{Summary and conclusions}
\label{summary}

We presented the optical nuclear spectra of 26 radio-sources from the
3C sample with a redshift $0.3 < z < 0.82$ (median redshift 0.51)
obtained from observations with VLT/MUSE integral field unit. The aim
of the present analysis is to derive an optical spectroscopic
classification of these sources based on the nuclear emission line
ratios, to compare the properties of the various sub-classes, and to
further explore their nature.

At these redshifts, the optical spectra do not include some of the
emission lines (e.g., the \Ha\ and [N~II] lines) that are generally used for
this classification. We then considered alternative diagnostic
diagrams (that we define as blue-DDs) including only emission lines
with a rest-frame wavelength of $\lambda \leq 5007$\AA\ (i.e., [Ne~V],
[O~II], [Ne~III], H$\beta$, and [O~III]). We showed that the blue-DDs
provide a robust spectral classification. We found, among the 26 izRGs
observed, six LEGs and 20 HEGs; eight sources of the latter class show
broad emission lines. No star-forming galaxies are present in the
sample. We also found that a diagnostic diagram based on the relative
strengths of UV lines only ([Ne~V], [O~II], and [Ne~III]) provides a
clear separation between the various classes: this result is
particularly important because: 1) it can be used for emission line
galaxies with redshift up to $z\sim1.5$ and 2) these line are very
close in wavelength, making their ratios robust against uncertainties
in the correction for dust absorption, a significant issue in this
spectral region.

The positive connection between line luminosity and radio power,
already found for lower redshift sources, continues in the izRGs.
However, the majority of these sources lie below the extrapolation of
the $L_{\rm[O~III]}$ - $L_{178}$ relation of the objects of these same
spectroscopic classes at low-z, that is, the slope of radio-line
correlation decreases at the highest power. It is likely that in more
luminous AGNs a higher fraction of nuclear ionizing photons escape
without being reprocessed by the ISM -- a similar effect to what has already
been found in QSOs.

We addressed the question of the link between the different
spectroscopic classes associated with FR~II radio galaxies. We found
that the fraction of LEGs does not vary significantly with increasing
redshift and that LEGs reach the highest level of radio power. The
similarity between the radio properties of HEGs and LEGs suggests that
the jet properties of the two RGs classes do not depend on the
accretion mode and on the structure of the accretion disk. This is
what is expected if the jet launching process in powerful
  radio-loud AGN is due to the extraction of the rotational energy of
the supermassive black hole. The accretion mode (and rate) instead
guides the dichotomy of the high energy (in the optical, UV, and X-ray
bands) nuclear properties.

A possible alternative is that HEGs and LEGs are connected by a
temporal sequence, due to a recurrent variation of the accretion
rate. However, this is disfavored when looking at the spectroscopic
properties of the EELR that are closely connected with those of the
NLR: none of the radio galaxies we observed with MUSE is caught in the
transition between the two classes, namely, with a HEG nucleus and a LEG
EELR (or viceversa). When considering the sample as a whole, this sets
a lower limit to the timescale for the duration of the different
activity phases of $\gtrsim 2\times 10^6$ years. This value is larger
than the time of survival of the radio hot spots, of order of $\sim
5\times 10^5$ years, after a drop in the jet power.

Apparently, the two classes of radio galaxies are the result of
different manifestations of accretion onto a supermassive black hole.
Nonetheless, both populations are able to produce relativistic jets
with similar properties.

\begin{acknowledgements}
  We acknowledge the contribution of the anonymous referee whose
  comments helped us to improve the clarity of this manuscript.
  B.B. acknowledges financial contribution from the agreement ASI-INAF
  I/037/12/0.  G.V. acknowledges support from ANID program FONDECYT
  Postdoctorado 3200802. A.M. acknowledges financial support from
  PRIN-MIUR contract 2017PH3WAT. S.B. and C.O. acknowledge support
  from the Natural Sciences and Engineering Research Council (NSERC)
  of Canada.  Based on observations made with ESO Telescopes at the La
  Silla Paranal Observatory under programme ID 0106.B-0564(A). This
  research has made use of the CIRADA cutout service at URL
  cutouts.cirada.ca, operated by the Canadian Initiative for Radio
  Astronomy Data Analysis (CIRADA). CIRADA is funded by a grant from
  the Canada Foundation for Innovation 2017 Innovation Fund (Project
  35999), as well as by the Provinces of Ontario, British Columbia,
  Alberta, Manitoba and Quebec, in collaboration with the National
  Research Council of Canada, the US National Radio Astronomy
  Observatory and Australia’s Commonwealth Scientific and Industrial
  Research Organisation.  This research has made use of the NASA/IPAC
  Extragalactic Database (NED), which is operated by the Jet
  Propulsion Laboratory, California Institute of Technology, under
  contract with the National Aeronautics and Space
  Administration. S. Baum and C. O'Dea are grateful to the Natural
  Sciences and Engineering Research Council (NSERC) of Canada.

\end{acknowledgements}

\bibliographystyle{./aa}
\bibliography{../../../bib} 

\begin{appendix}

\section{Nuclear spectra.}
\label{spectra}  
We present the nuclear spectra of the 26 izRGs obtained from the MUSE
observations, focusing on the spectral regions where the emission lines
of interest are located.

\begin{figure*}  
\includegraphics[width=0.49\textwidth,angle=180]{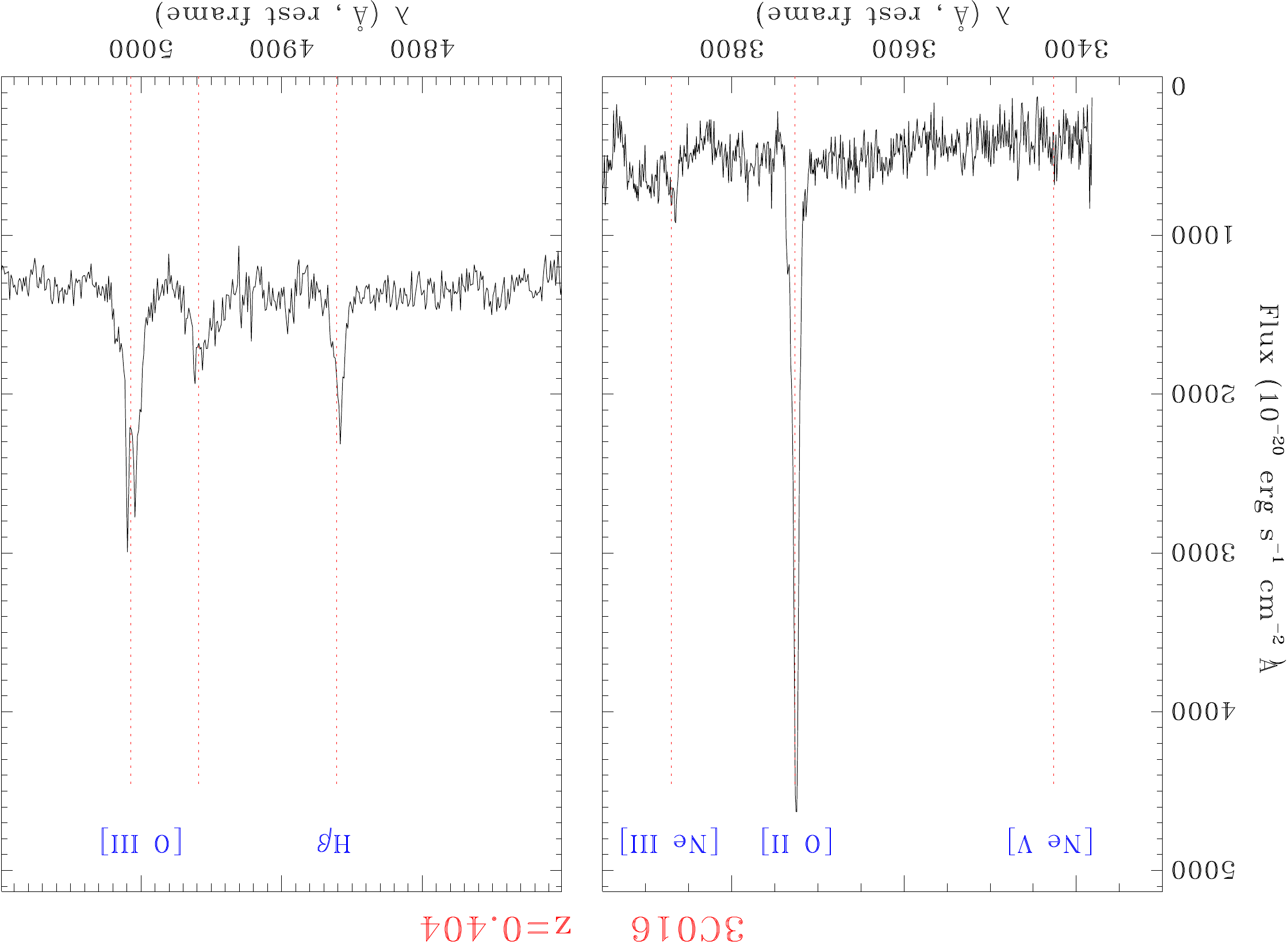}
\includegraphics[width=0.49\textwidth,angle=180]{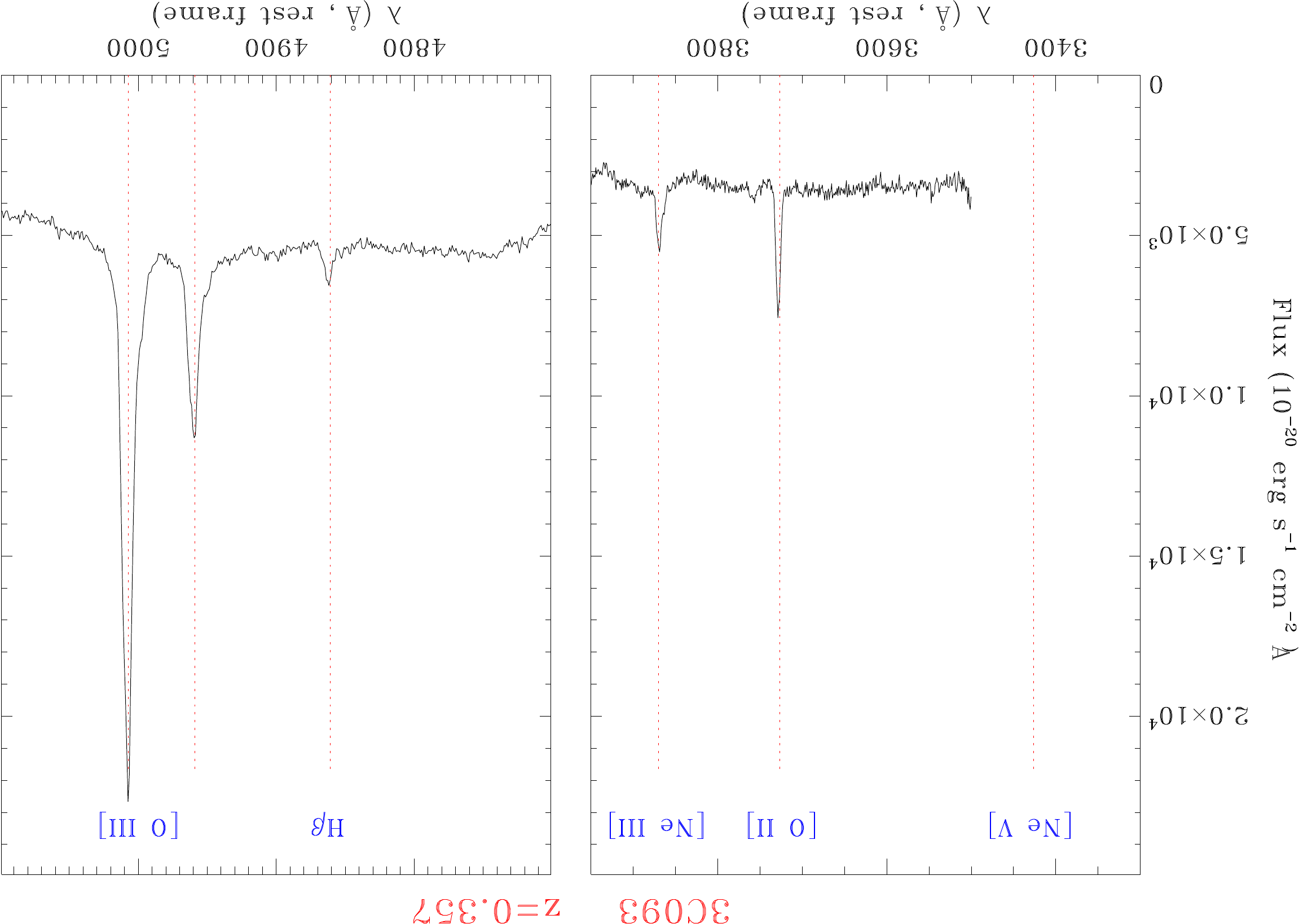}
\includegraphics[width=0.49\textwidth,angle=180]{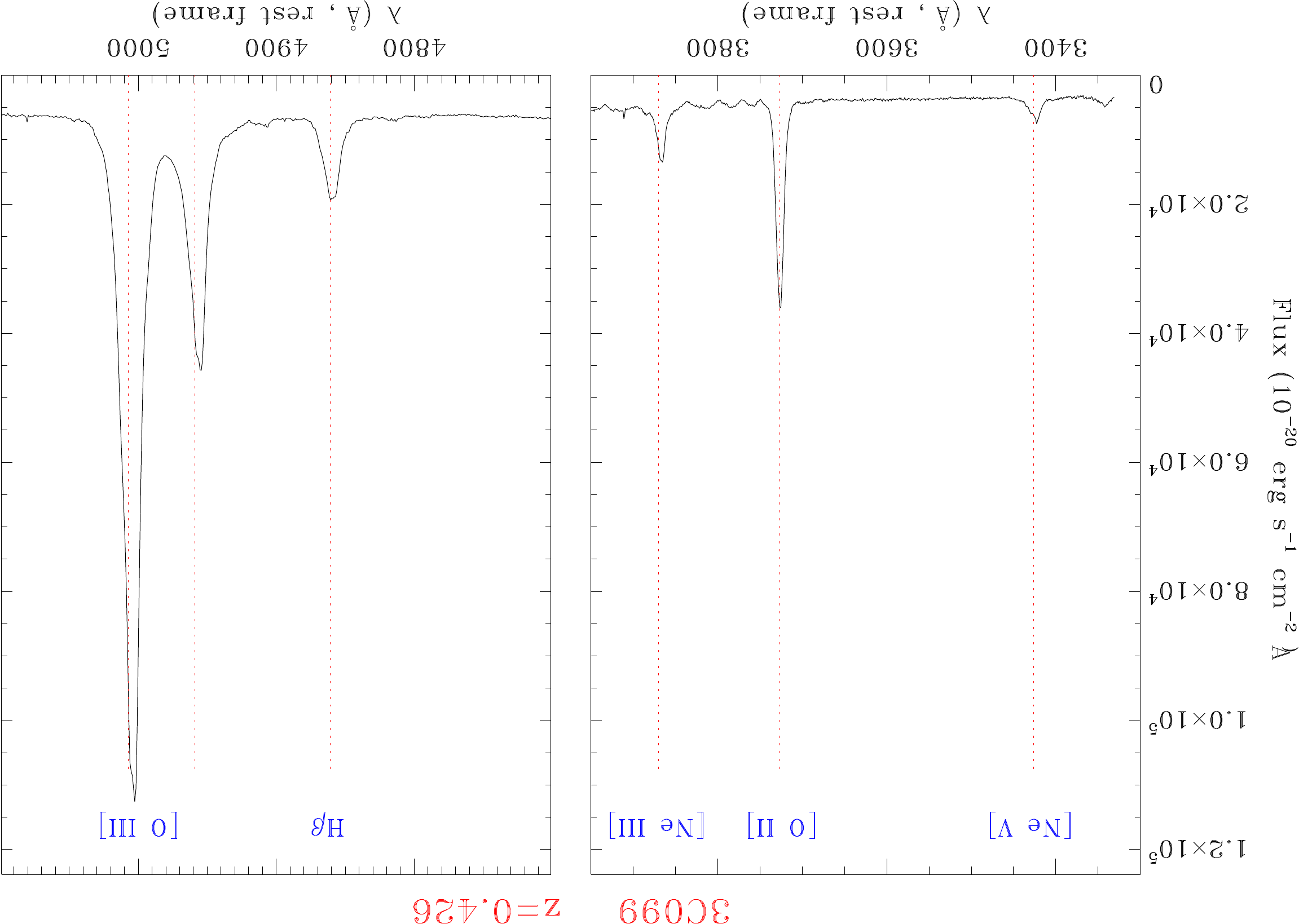}
\includegraphics[width=0.49\textwidth,angle=180]{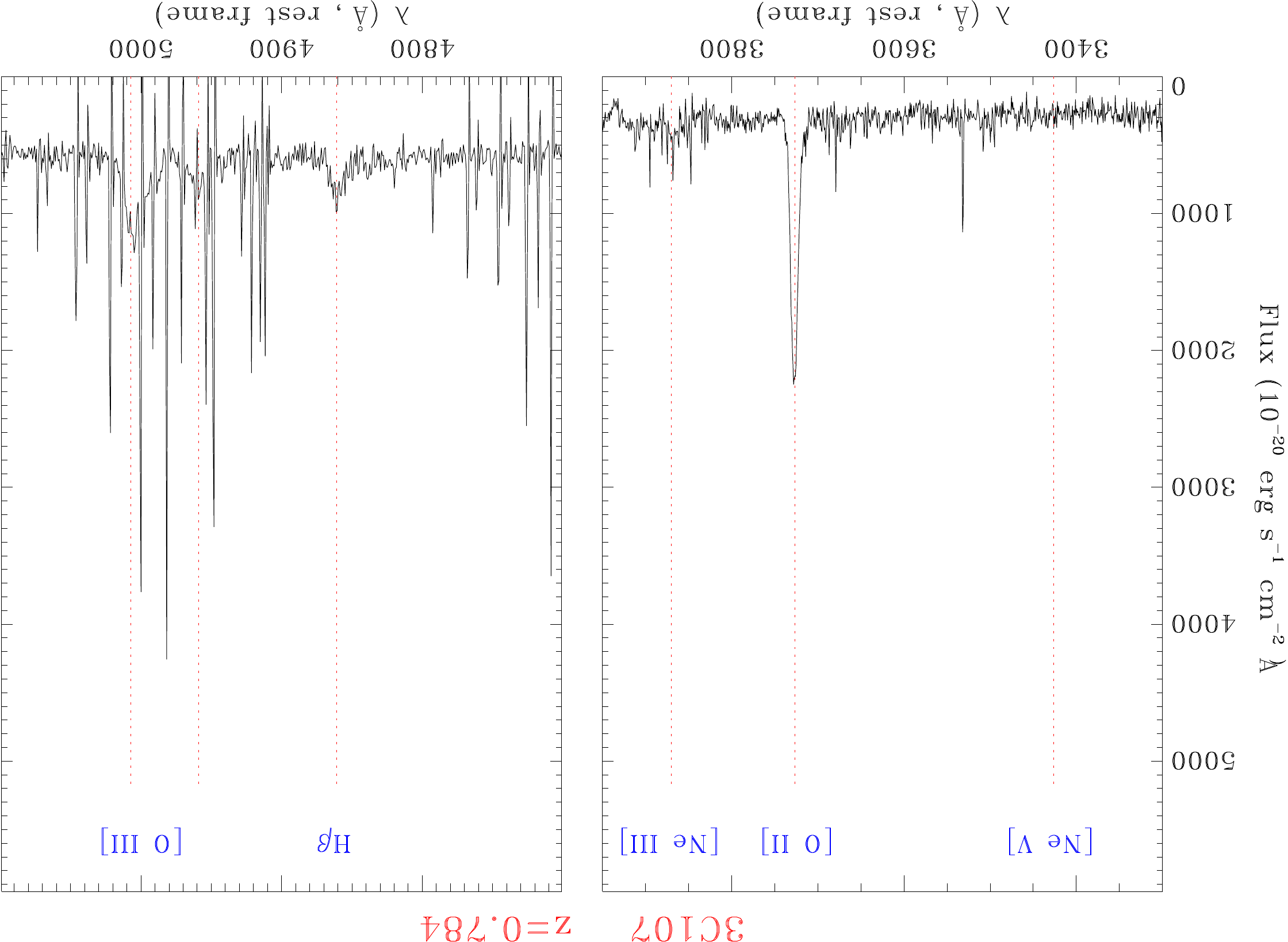}
\includegraphics[width=0.49\textwidth,angle=180]{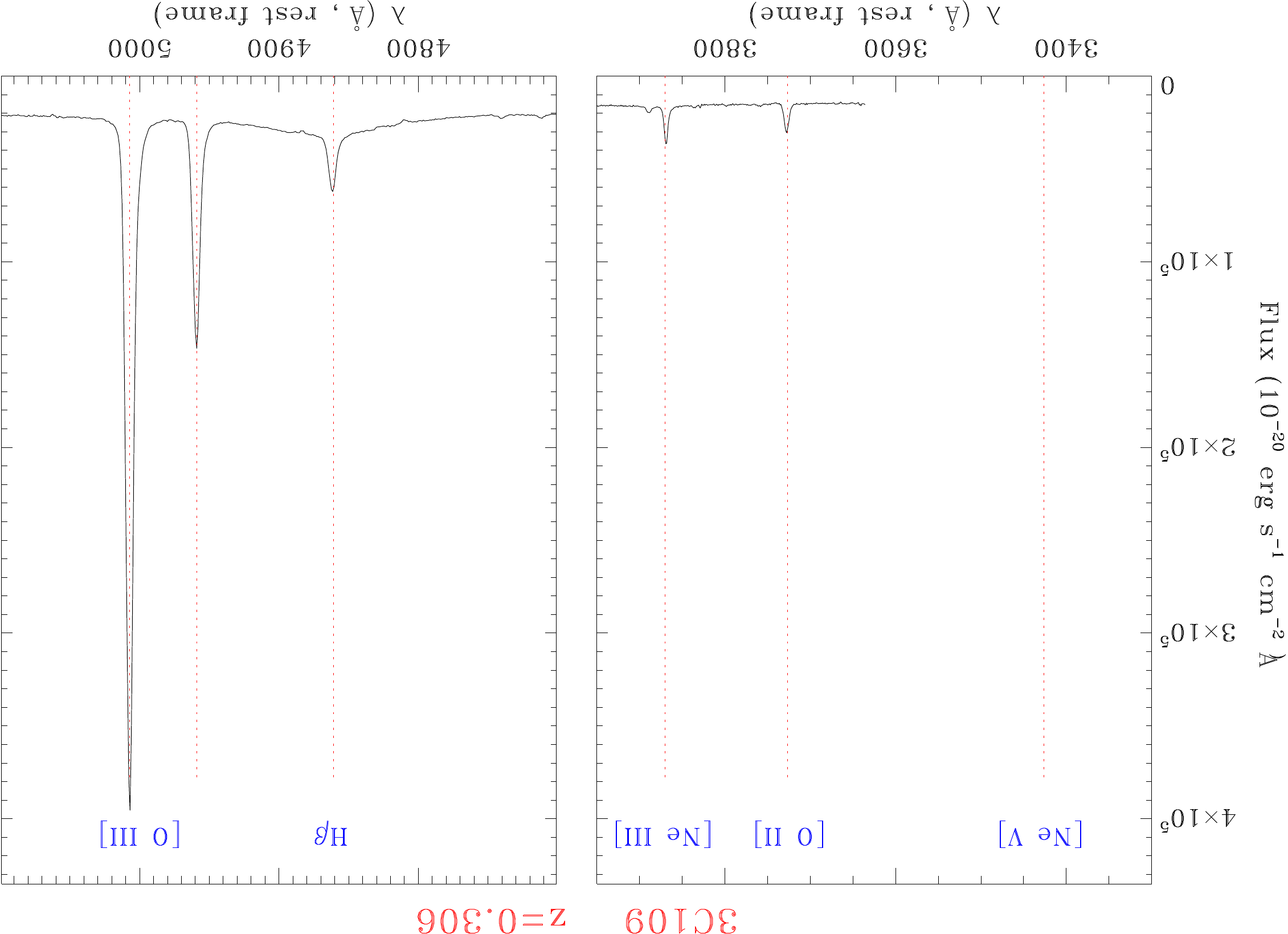}
\includegraphics[width=0.49\textwidth,angle=180]{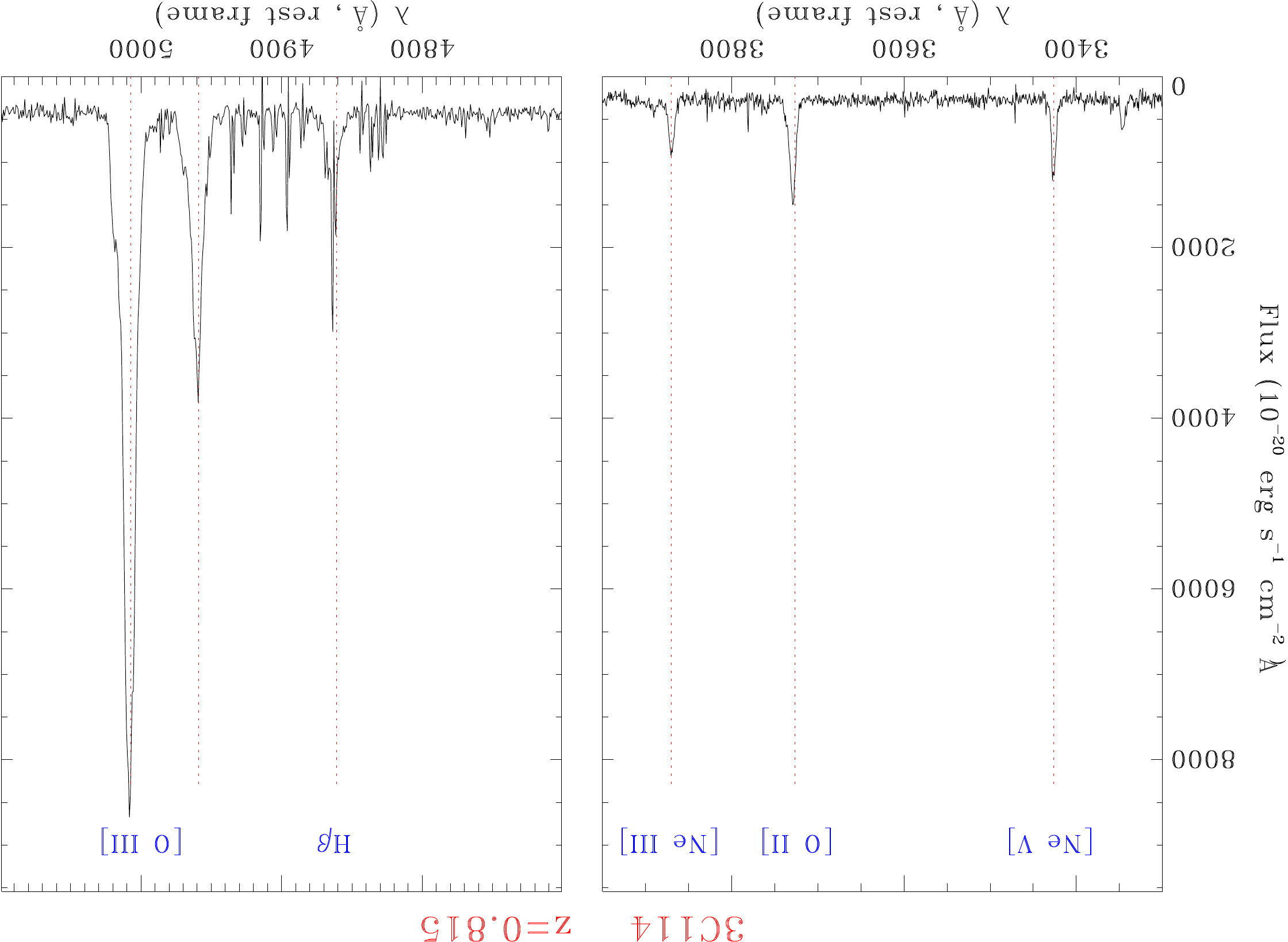}
\caption{Nuclear spectra of the 26 3C radio galaxies at 0.3 $<$ z $<$ 0.82.}
\label{blue}
\end{figure*}

\begin{figure*}  
\addtocounter{figure}{-1}
\includegraphics[width=0.49\textwidth,angle=180]{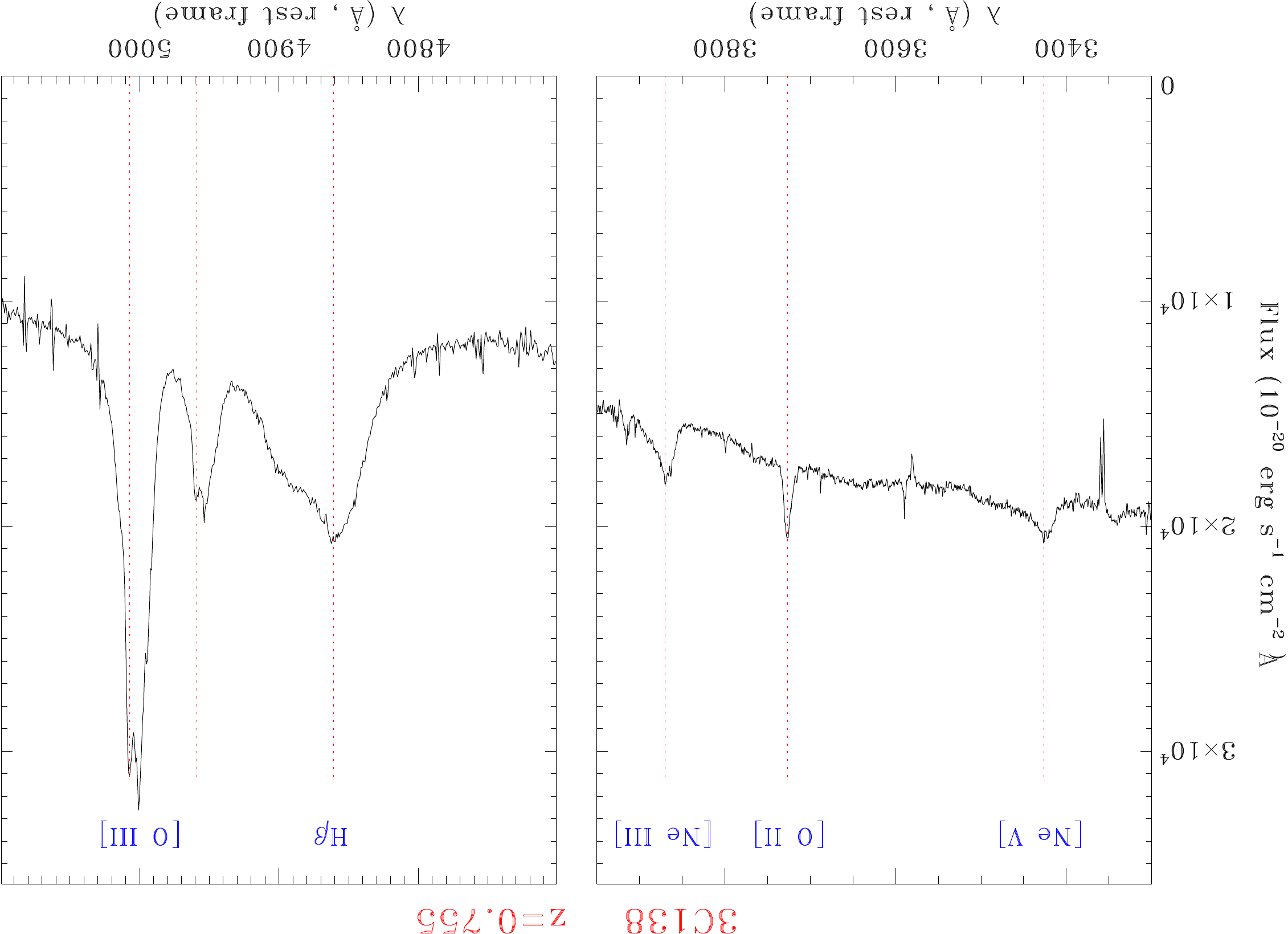}
\includegraphics[width=0.49\textwidth,angle=180]{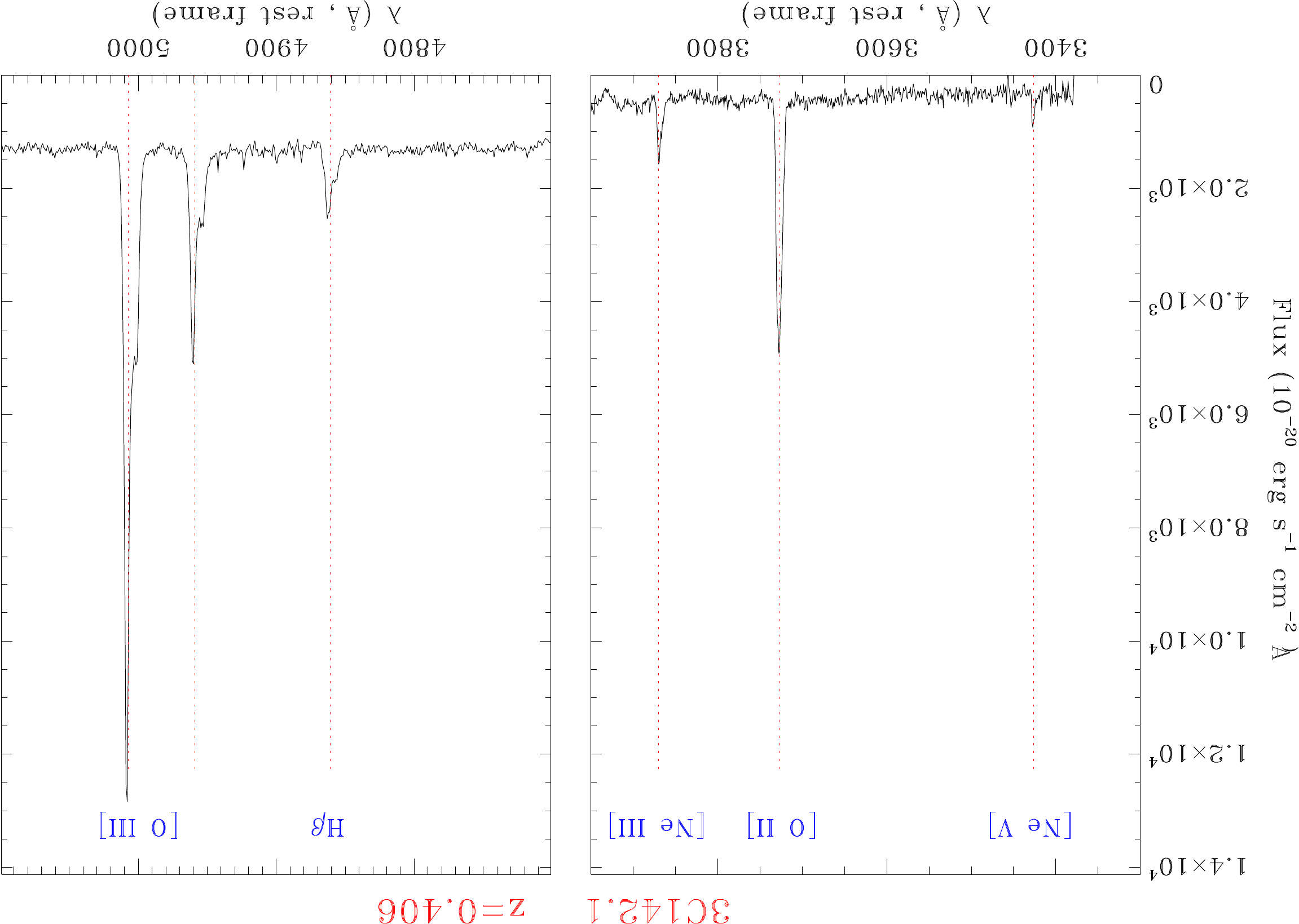}
\includegraphics[width=0.49\textwidth,angle=180]{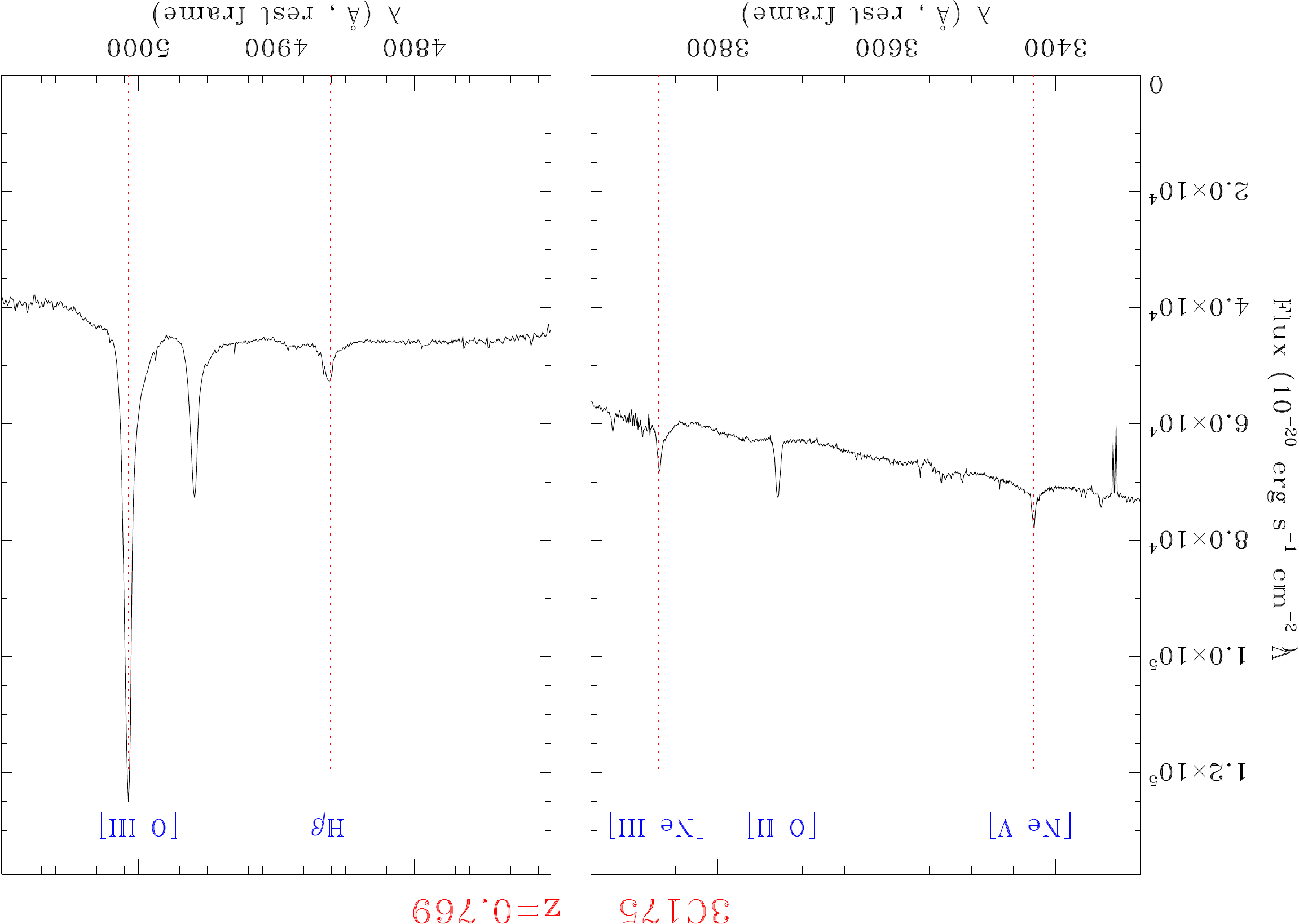}
\includegraphics[width=0.49\textwidth,angle=180]{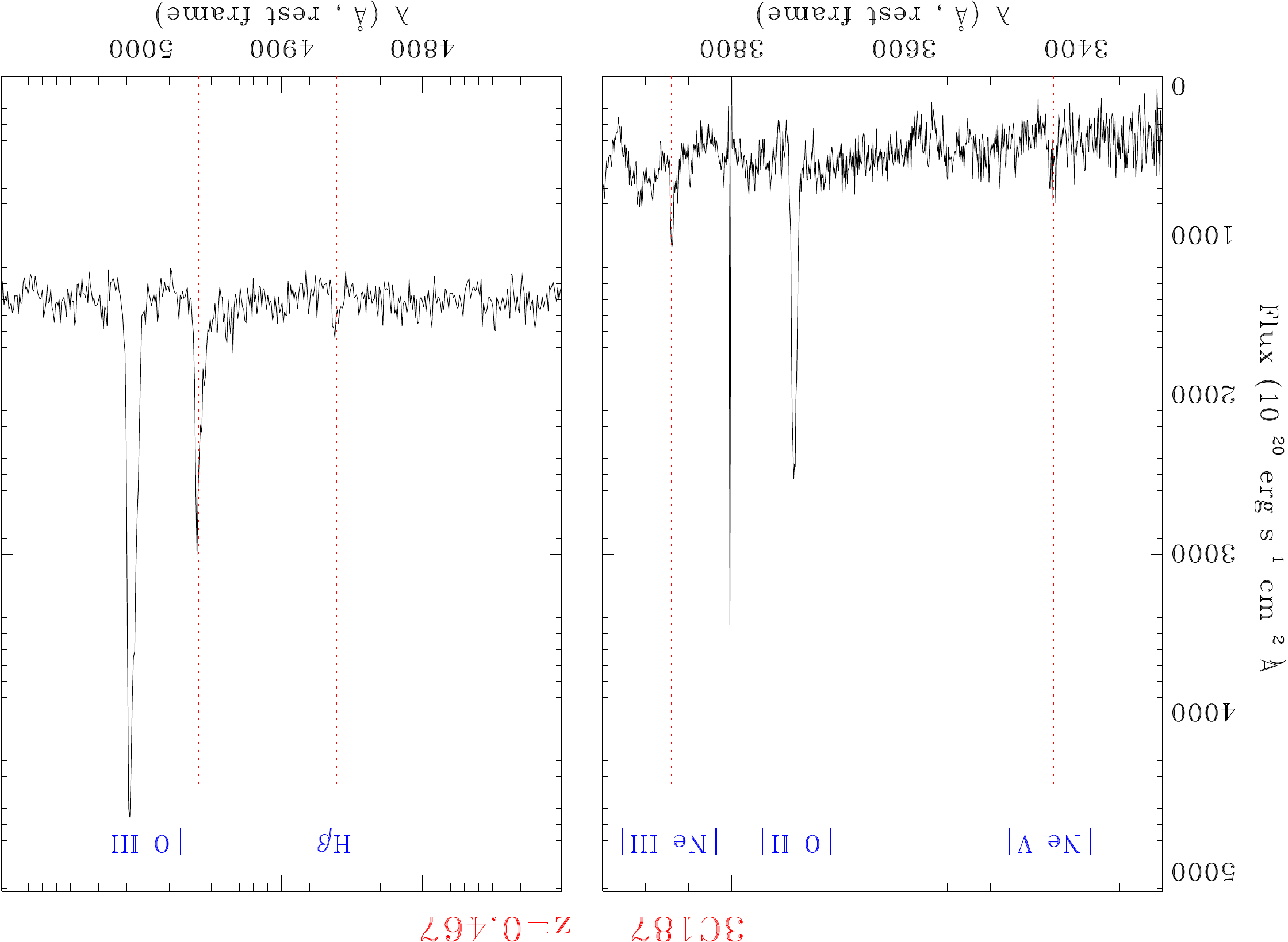}
\includegraphics[width=0.49\textwidth,angle=180]{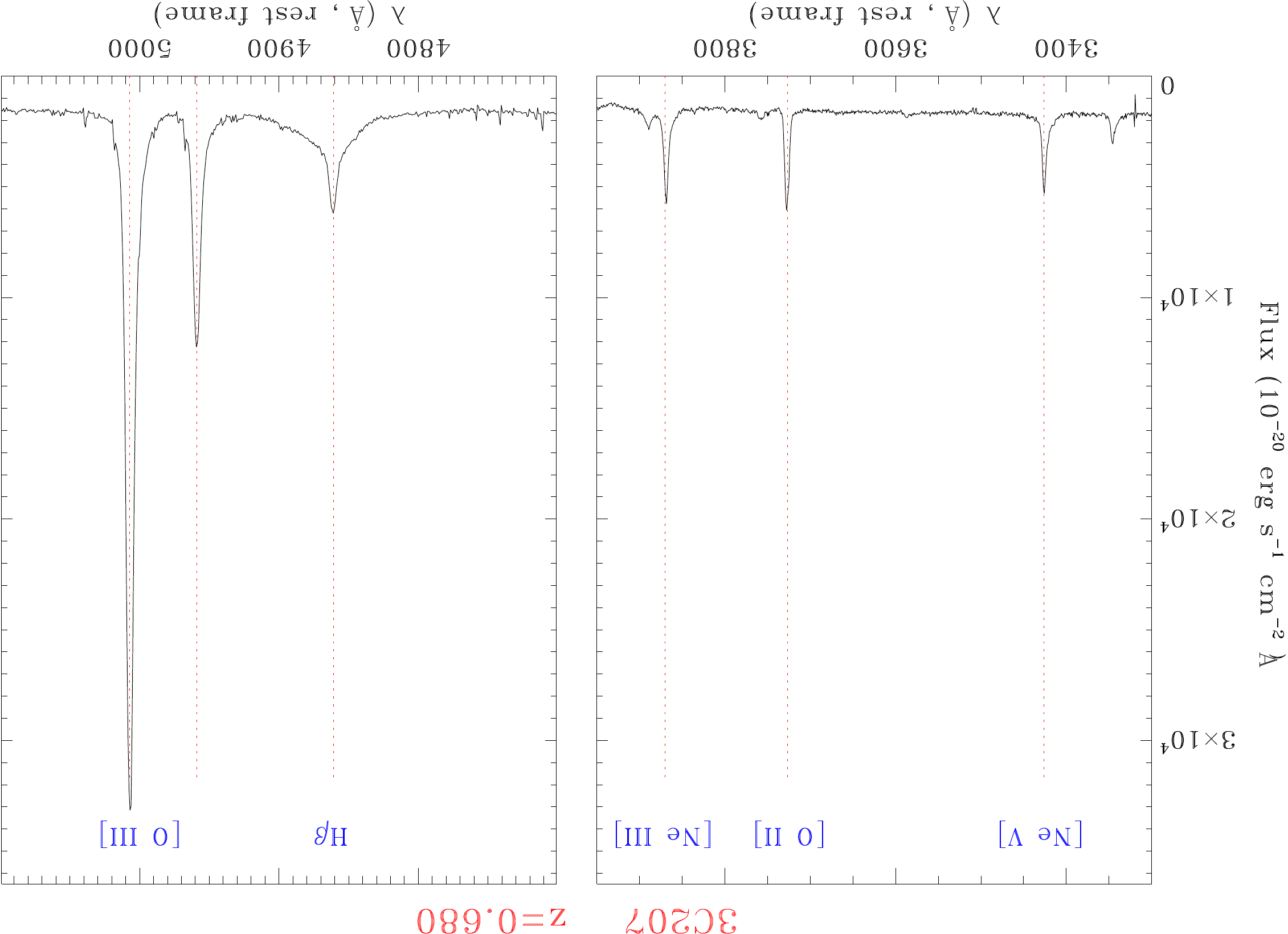}
\includegraphics[width=0.49\textwidth,angle=180]{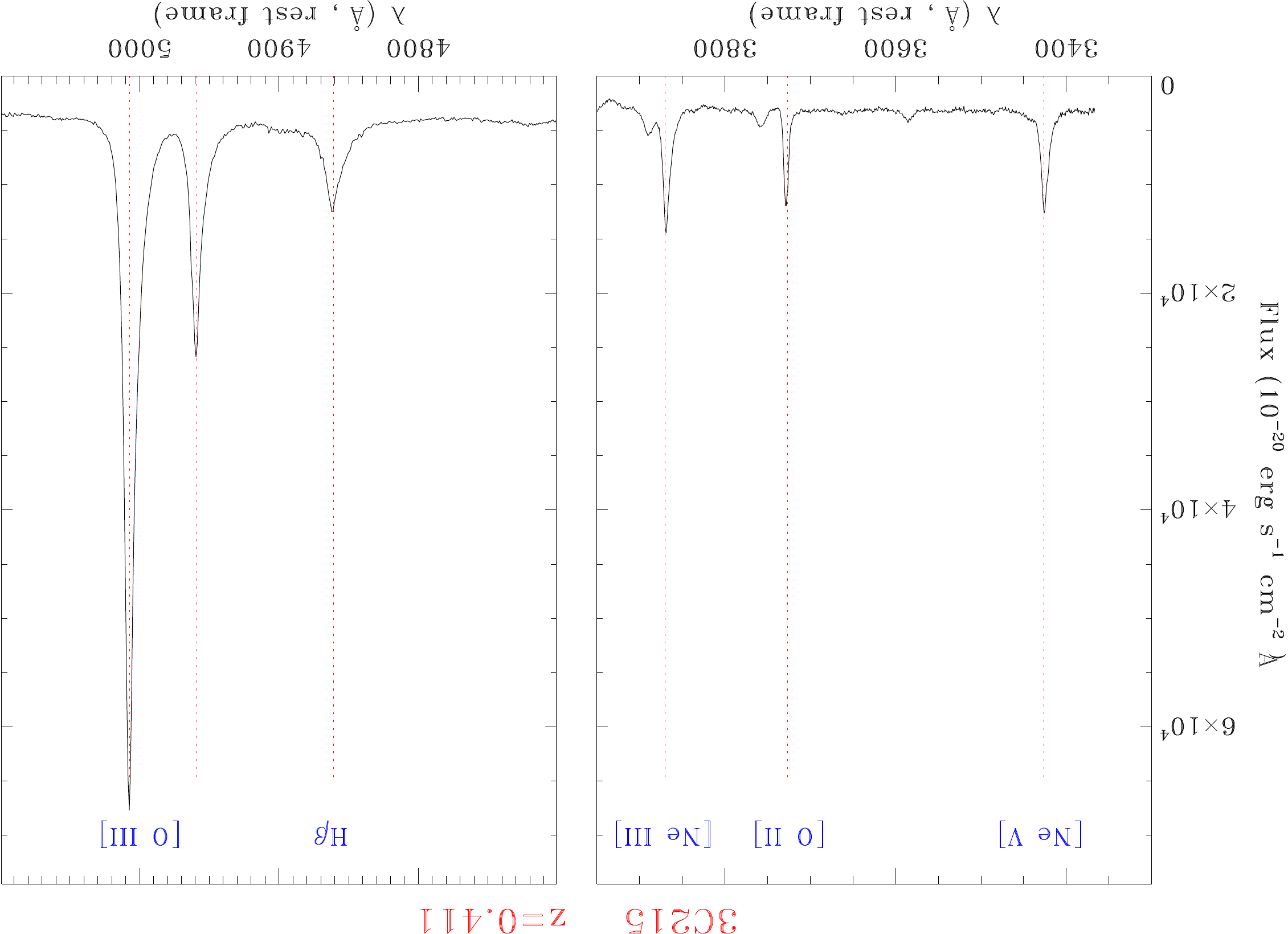}
\caption{- continued.}
\end{figure*}

\begin{figure*}
\addtocounter{figure}{-1}
\includegraphics[width=0.49\textwidth,angle=180]{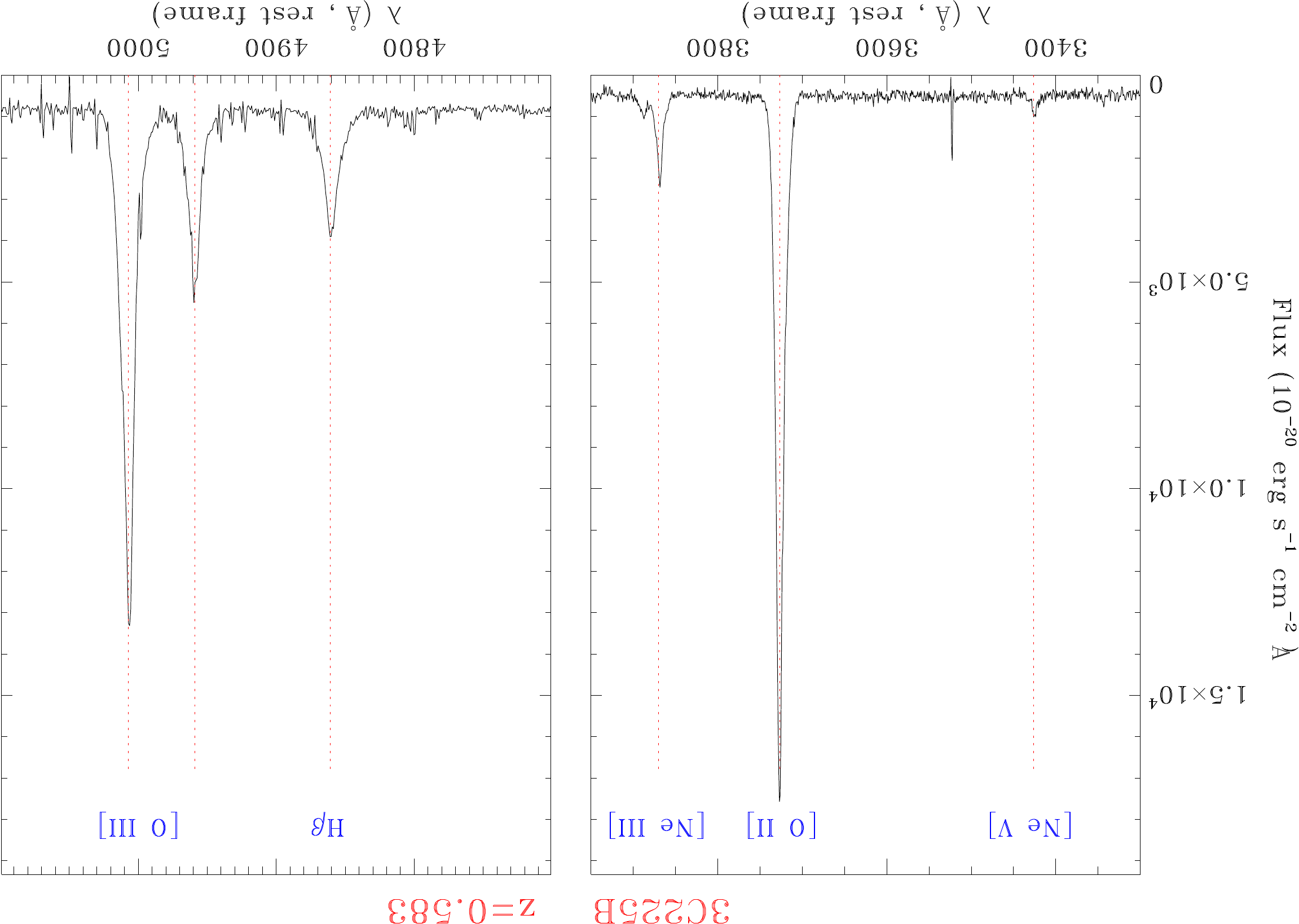}
\includegraphics[width=0.49\textwidth,angle=180]{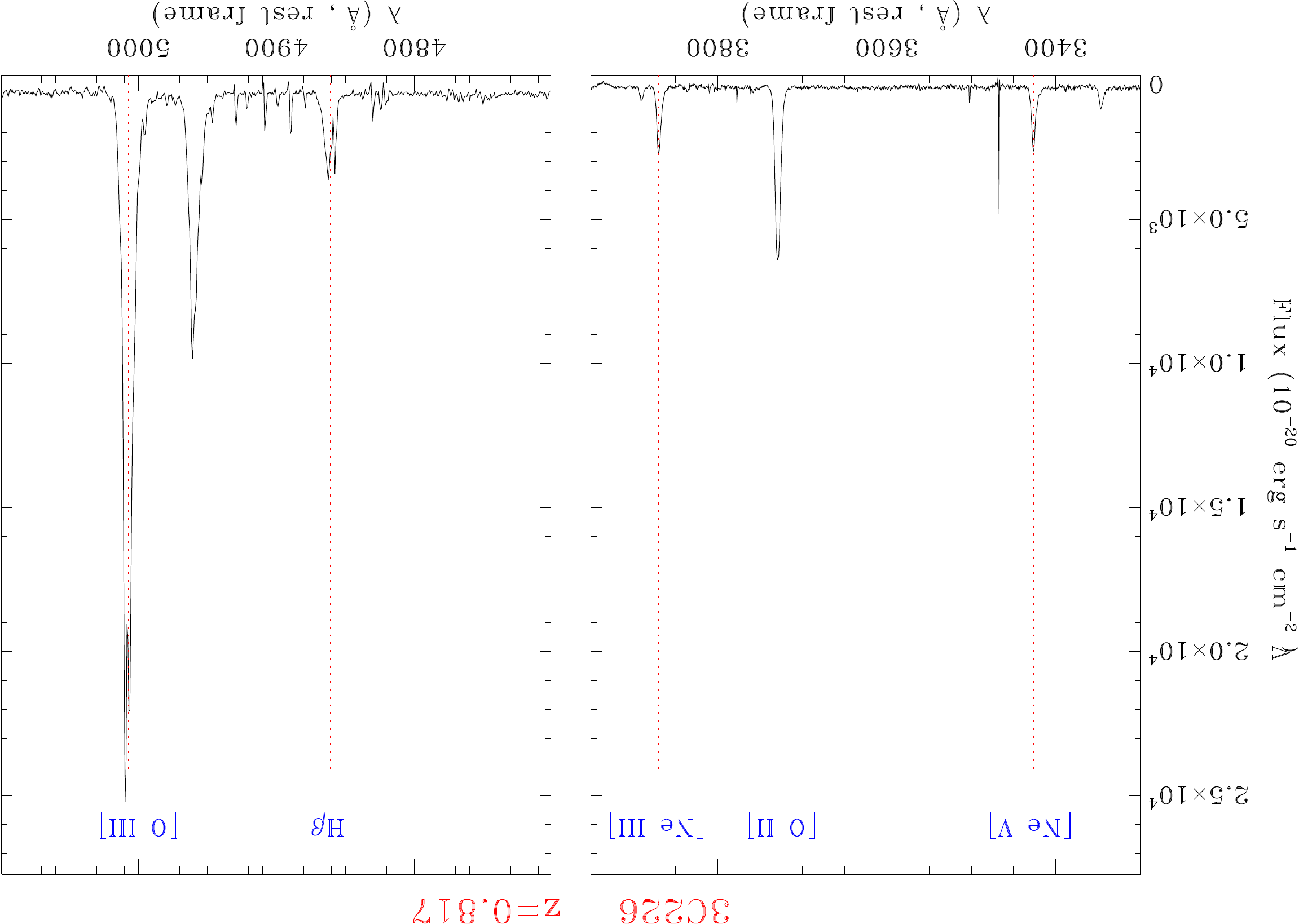}
\includegraphics[width=0.49\textwidth,angle=180]{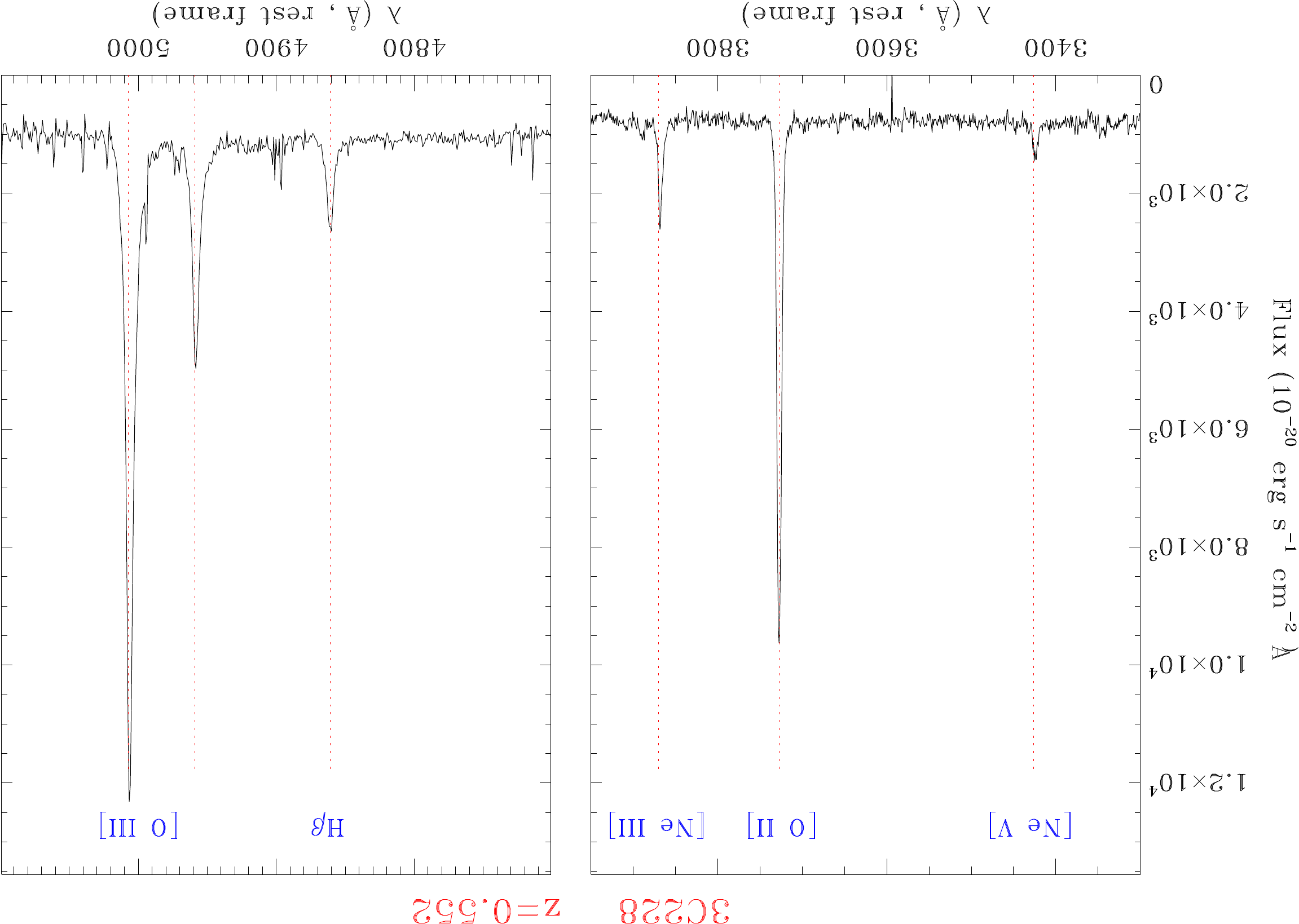}
\includegraphics[width=0.49\textwidth,angle=180]{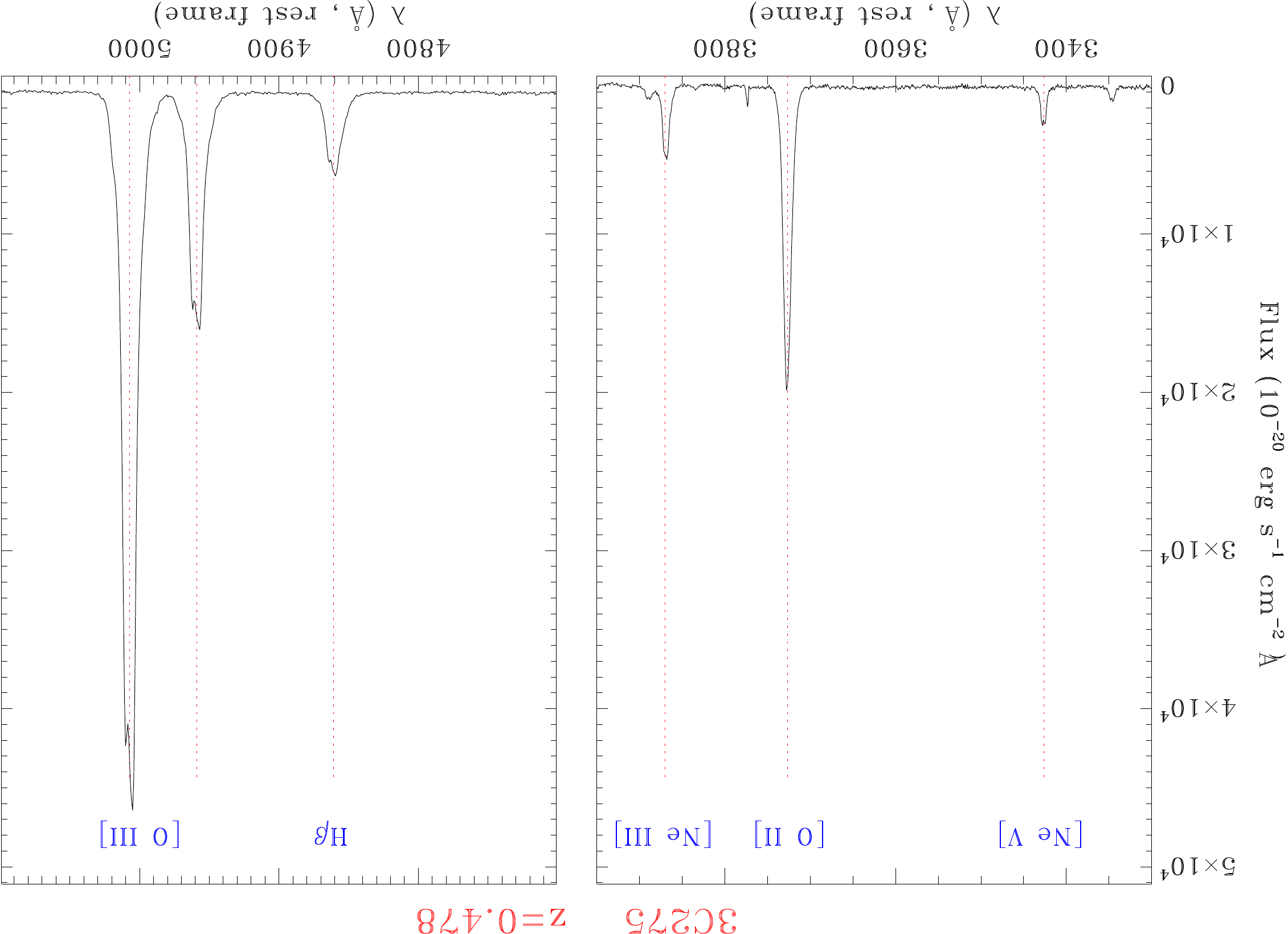}
\includegraphics[width=0.49\textwidth,angle=180]{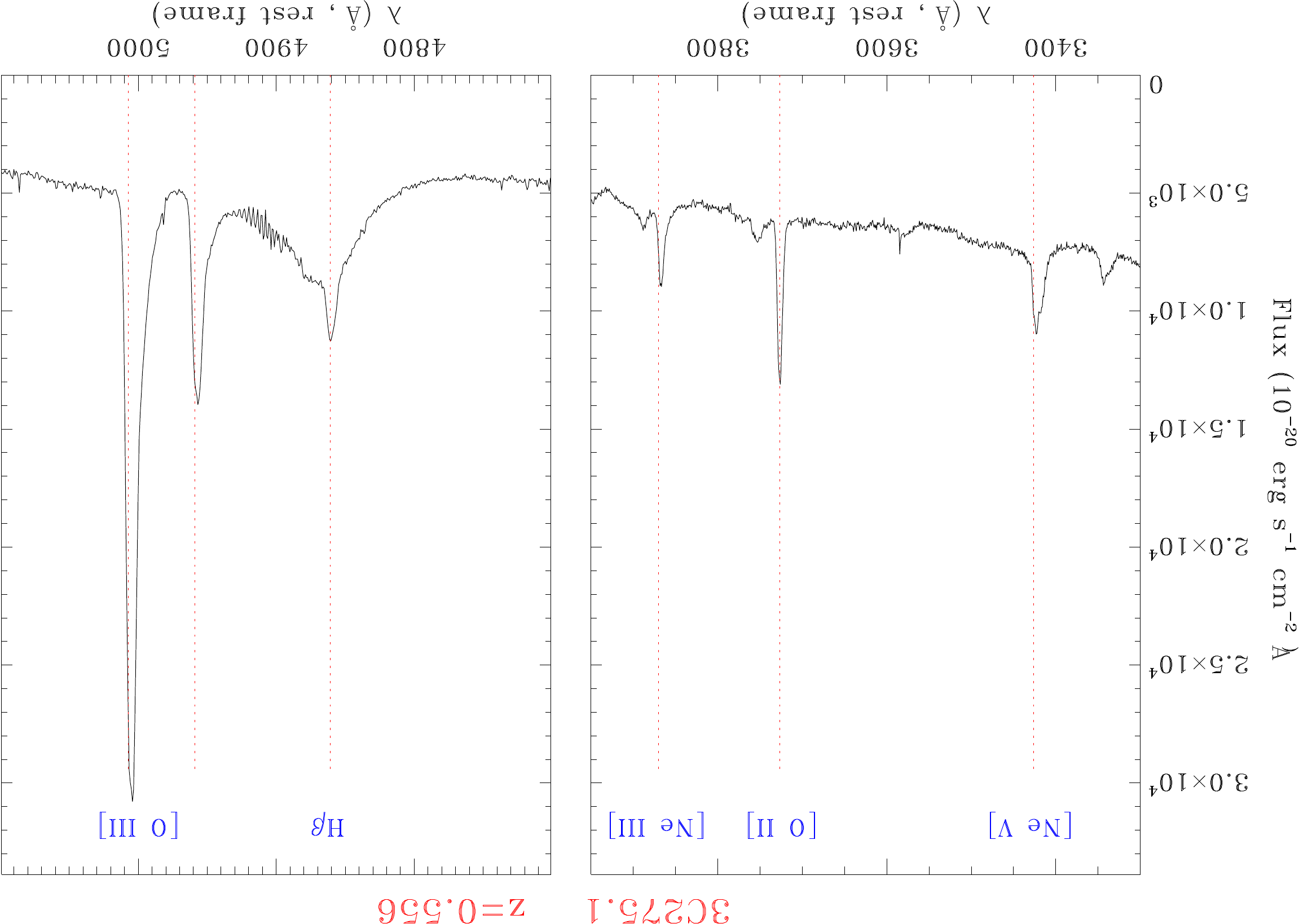}
\includegraphics[width=0.49\textwidth,angle=180]{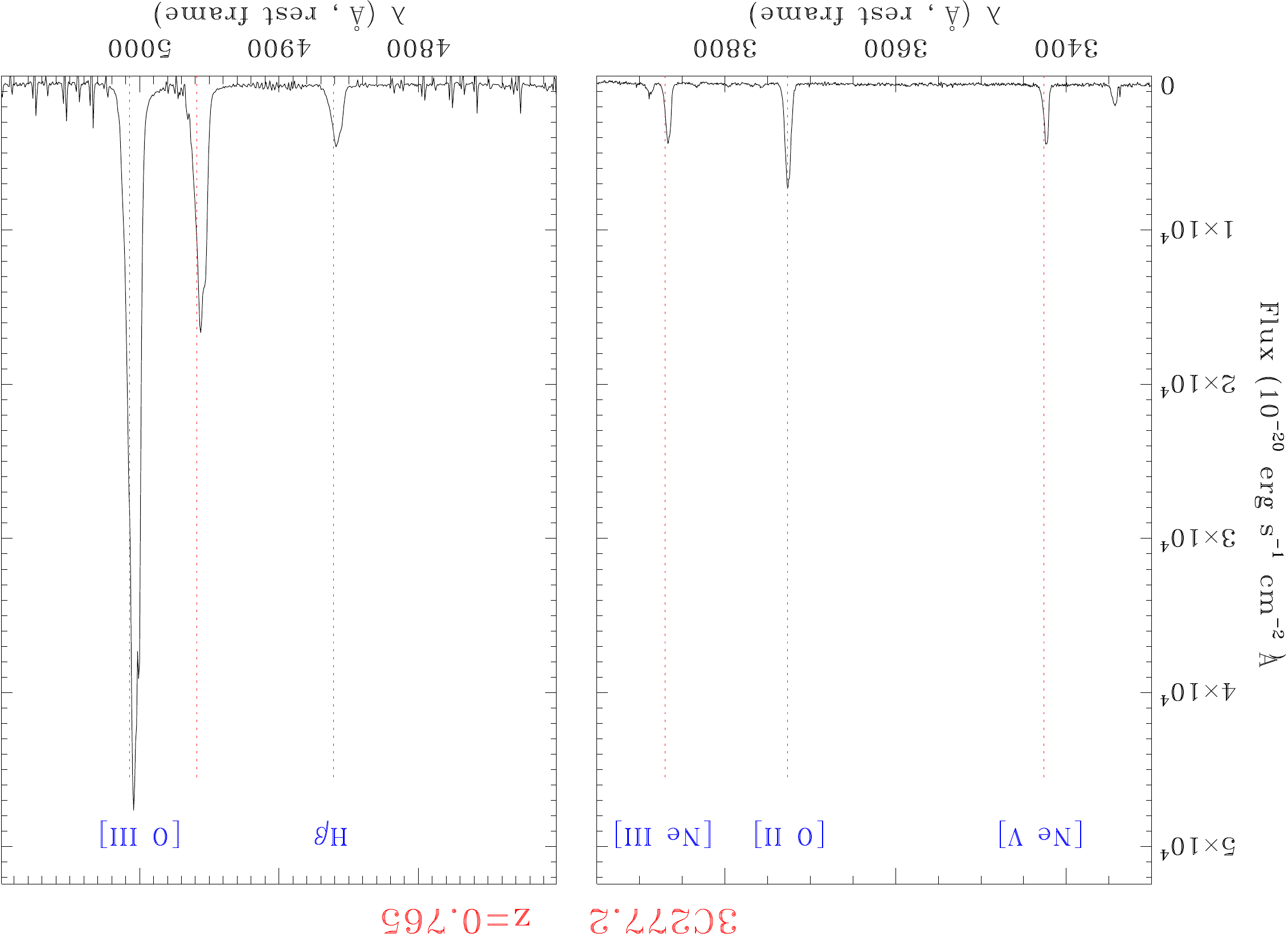}
\caption{- continued.}
\end{figure*}

\begin{figure*}
\addtocounter{figure}{-1}
\includegraphics[width=0.49\textwidth,angle=180]{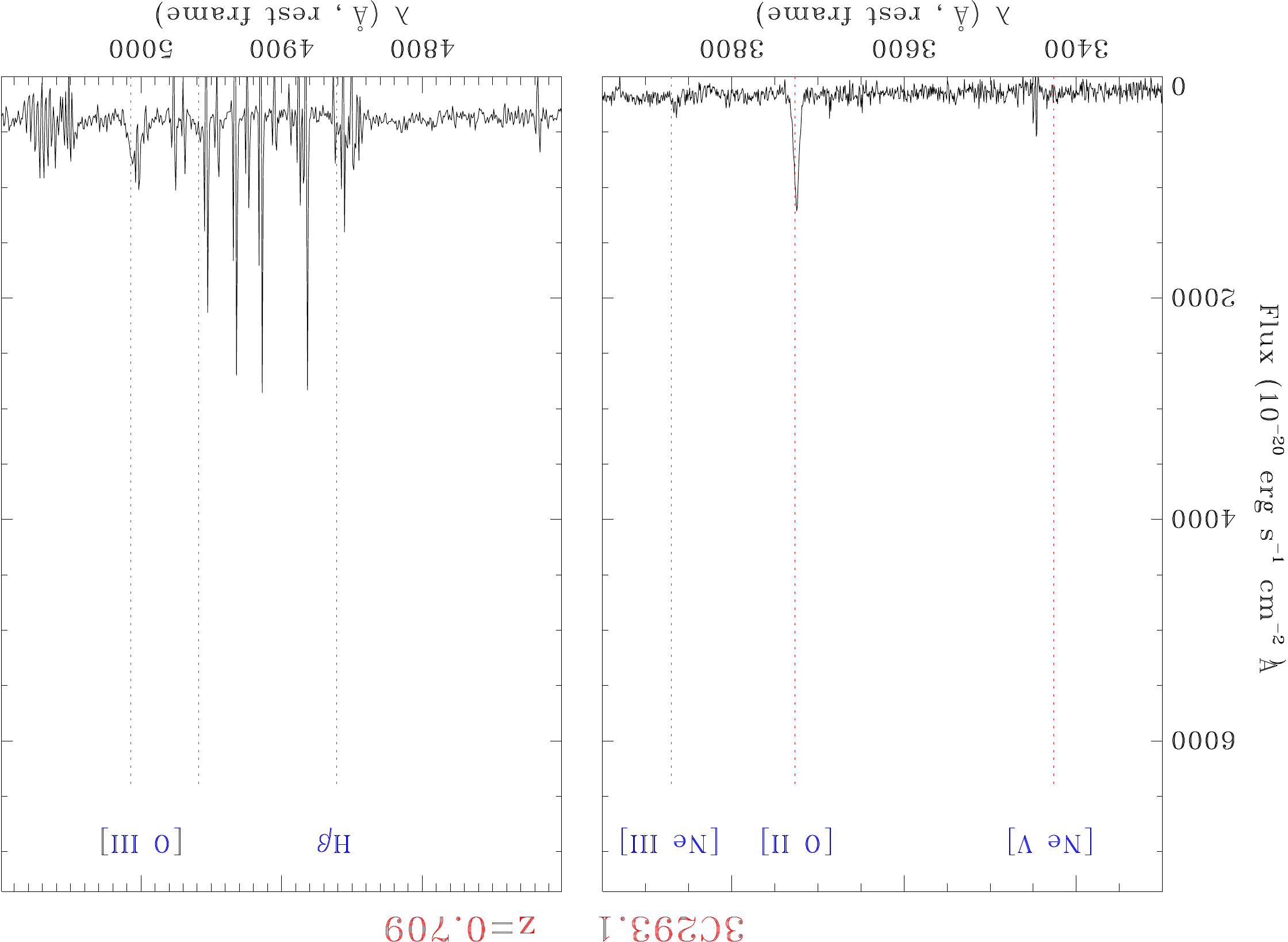}
\includegraphics[width=0.49\textwidth,angle=180]{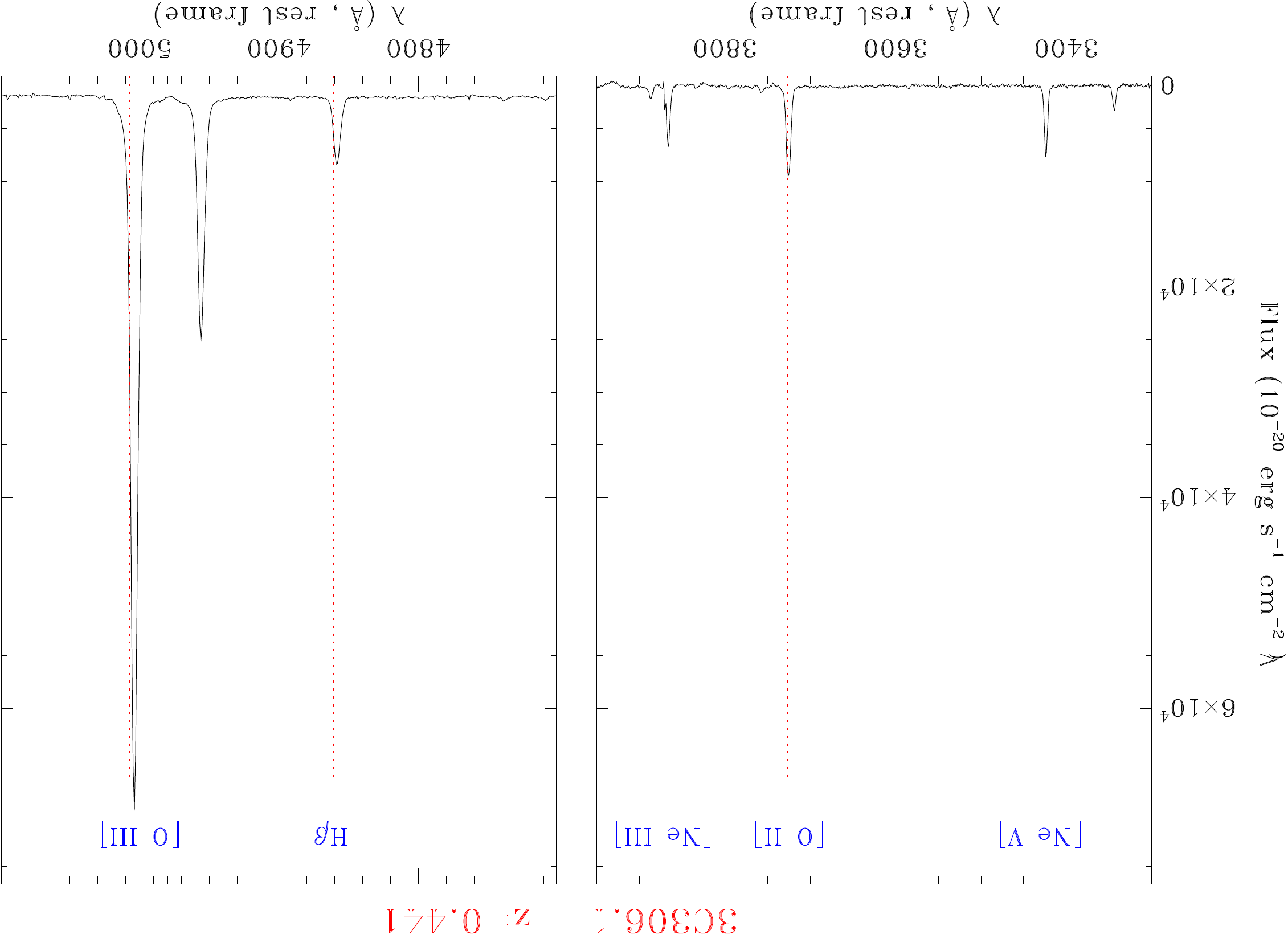}
\includegraphics[width=0.49\textwidth,angle=180]{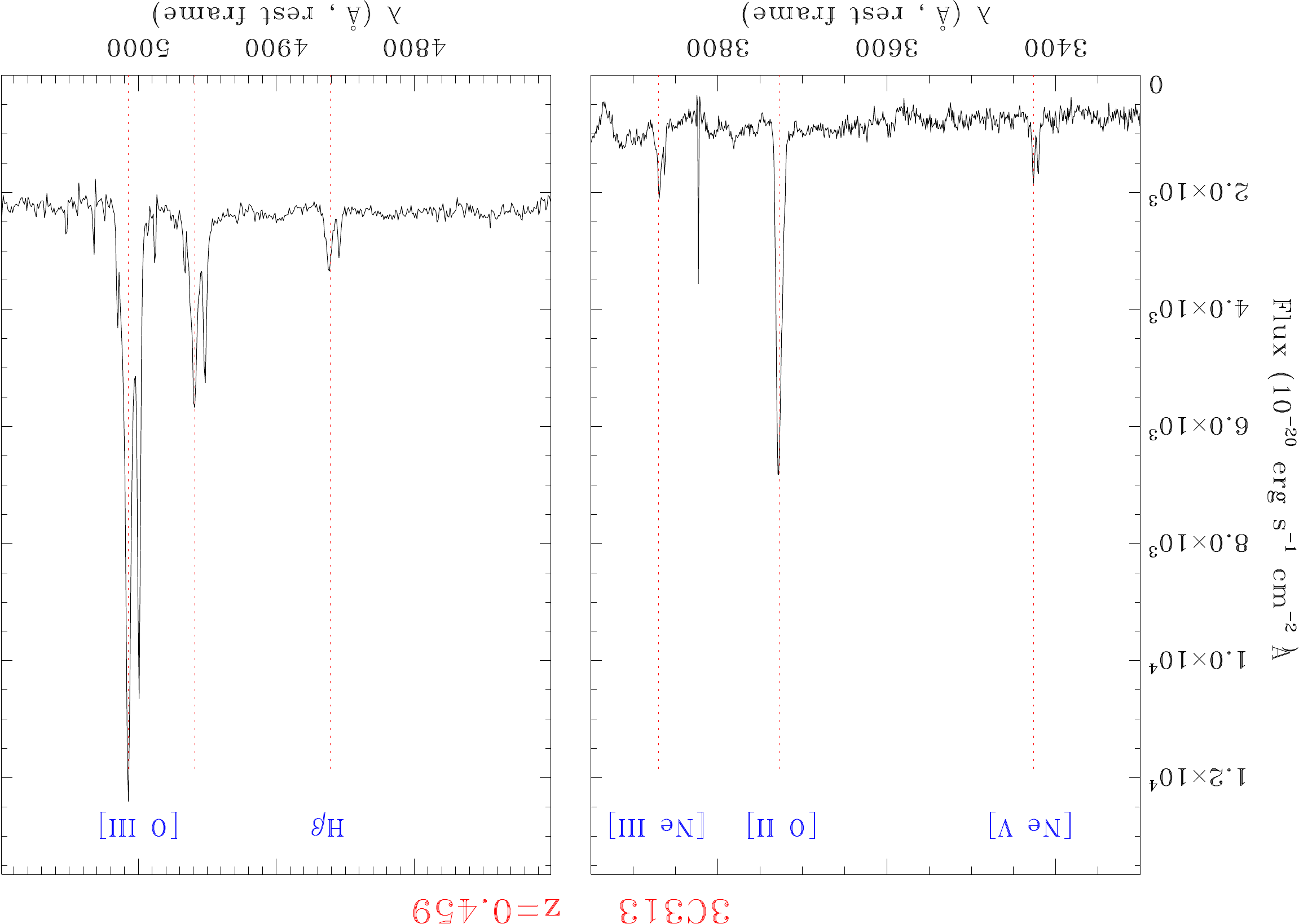}
\includegraphics[width=0.49\textwidth,angle=180]{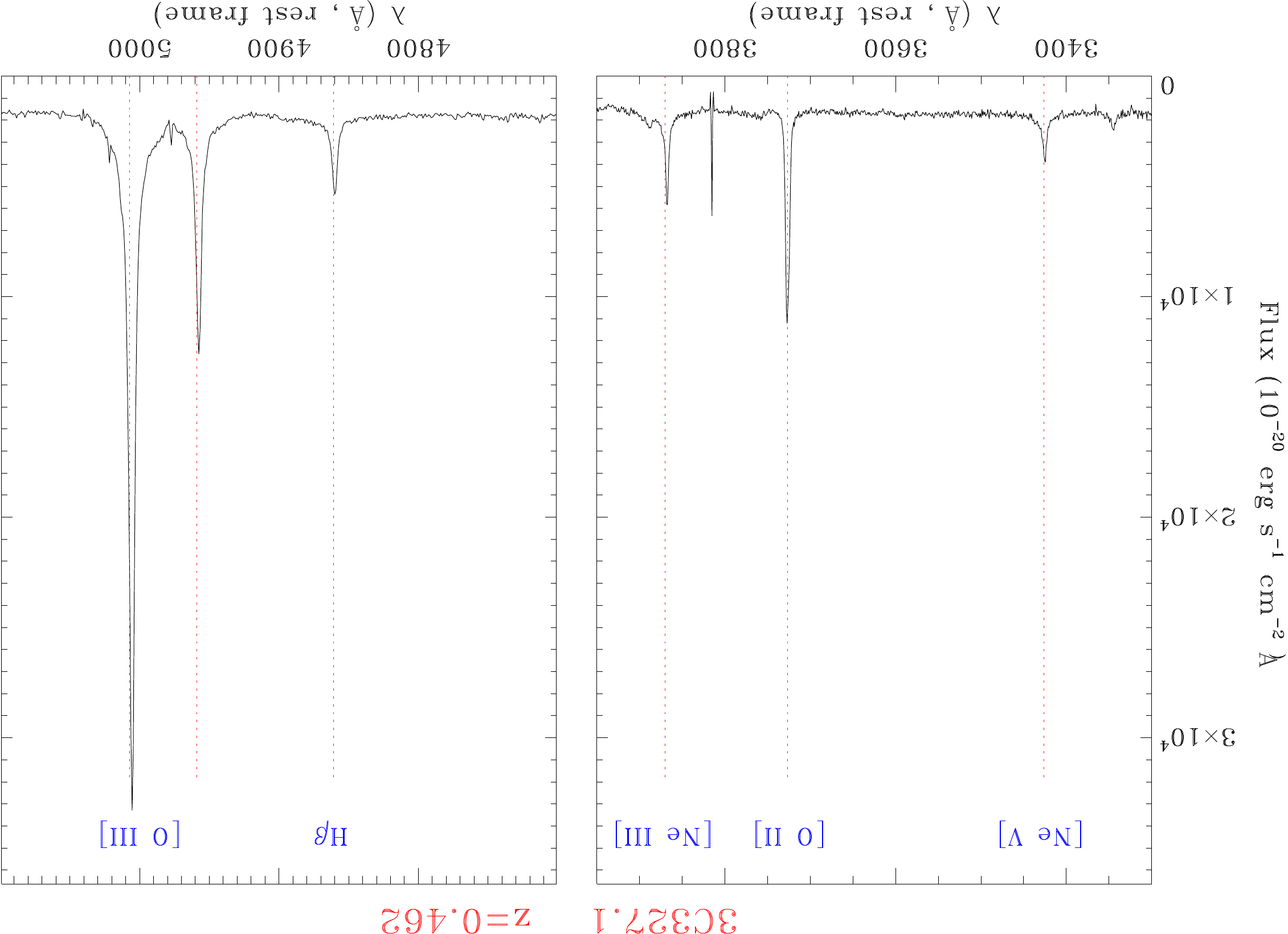}
\includegraphics[width=0.49\textwidth,angle=180]{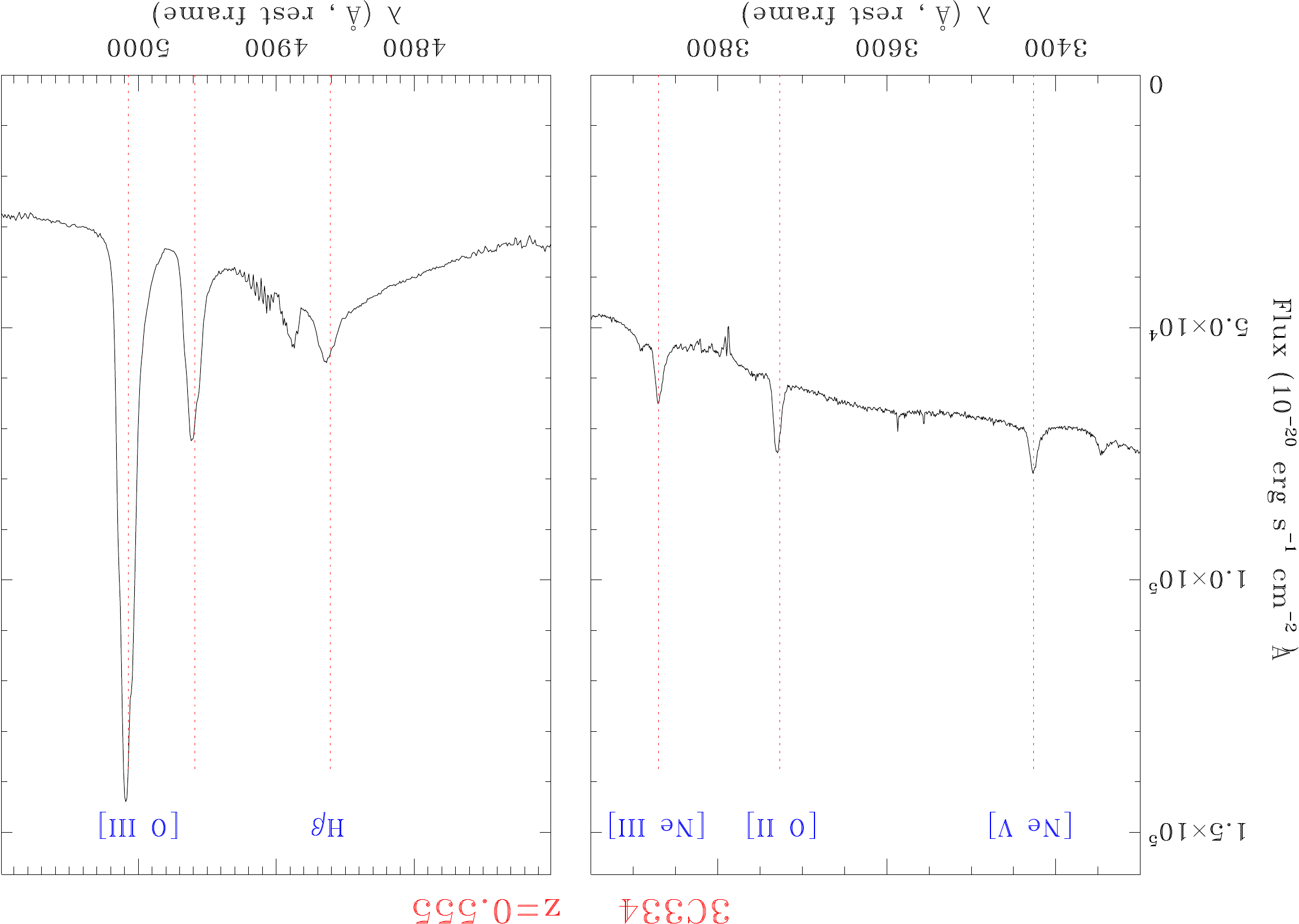}
\includegraphics[width=0.49\textwidth,angle=180]{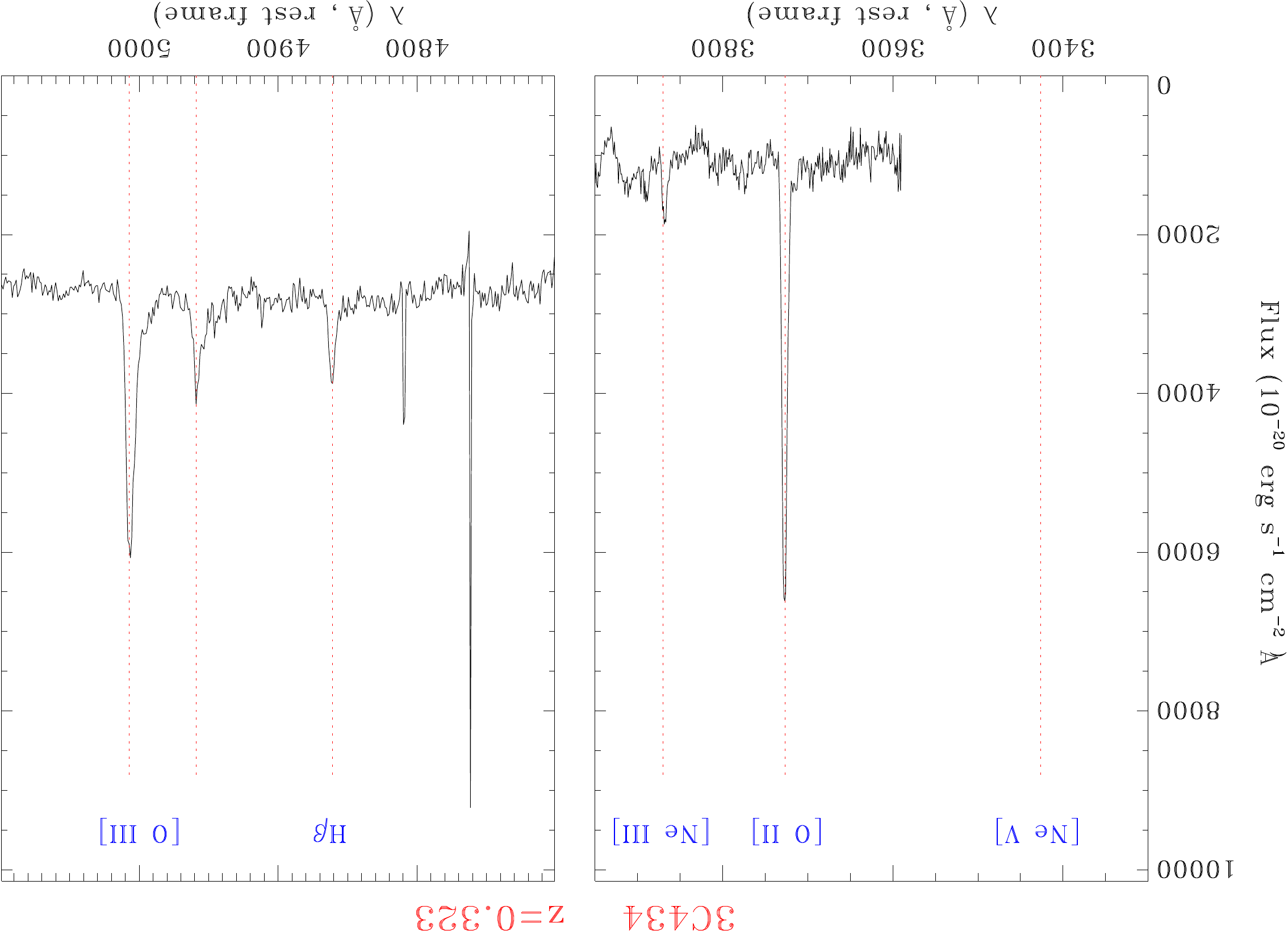}
\caption{- continued.}
\end{figure*}

\begin{figure*}
\addtocounter{figure}{-1}
\includegraphics[width=0.49\textwidth,angle=180]{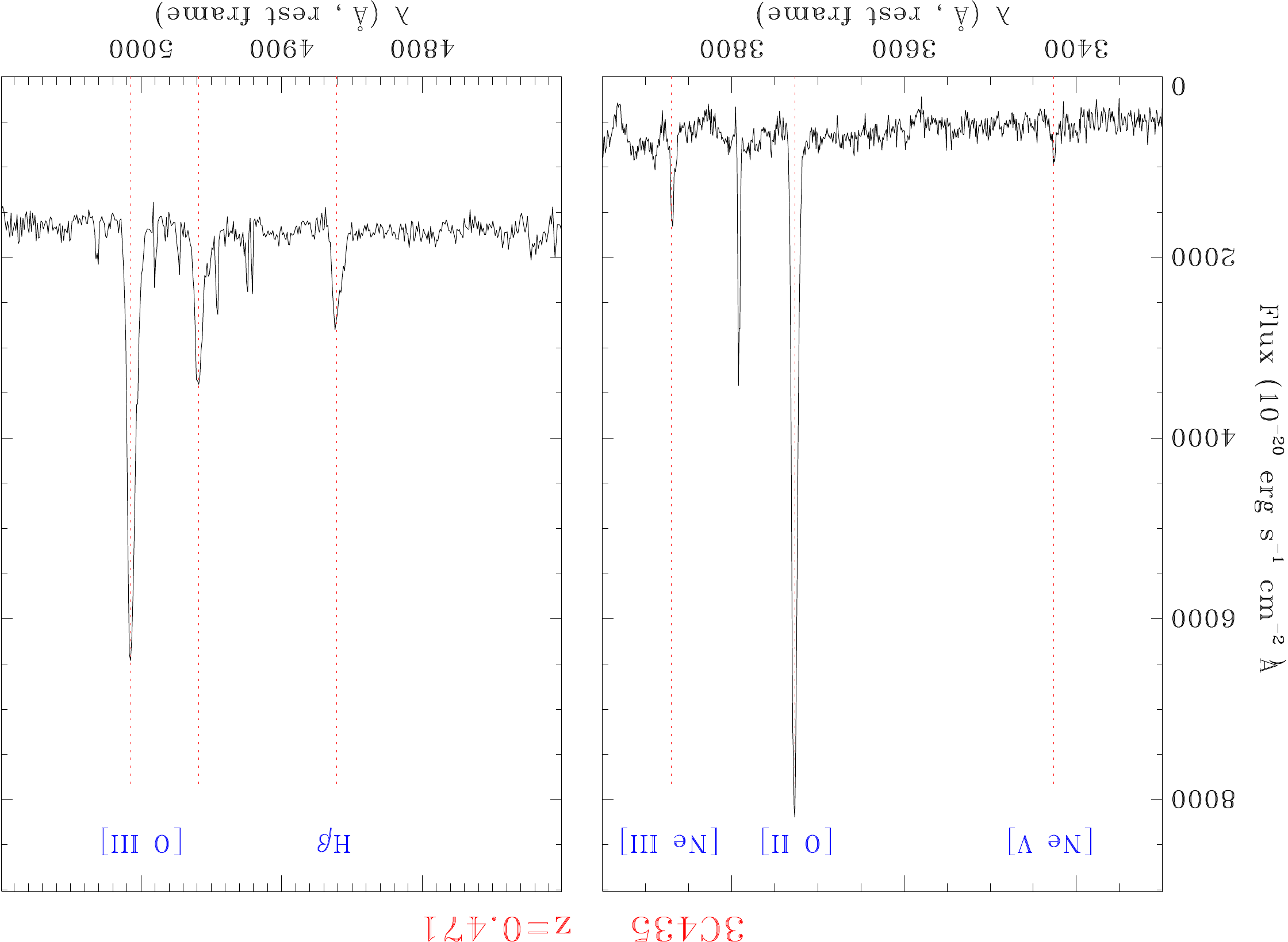}
\includegraphics[width=0.49\textwidth,angle=180]{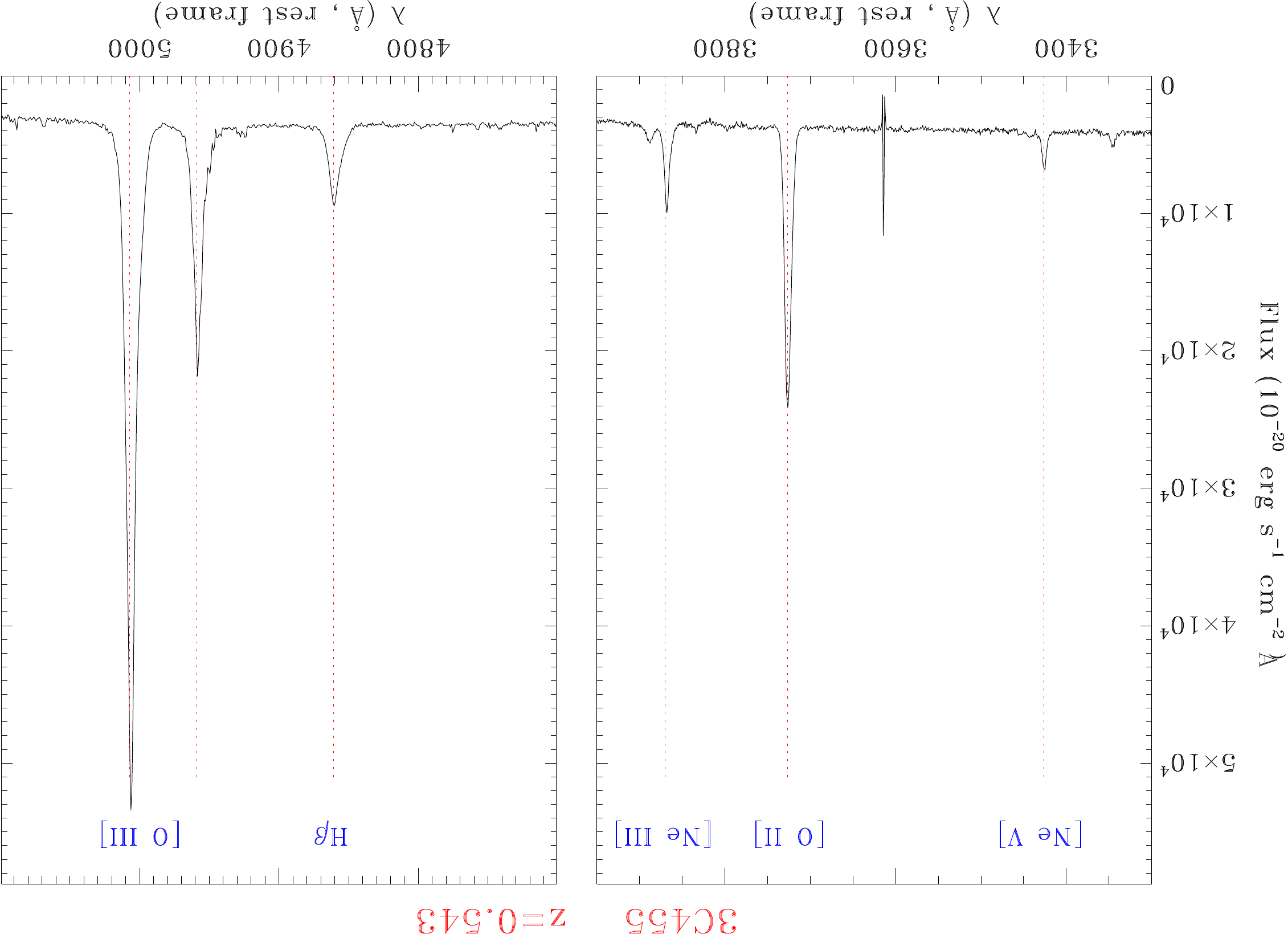}
\caption{- continued.}
\end{figure*}

\newpage
\clearpage
\section{Red portion of the nuclear spectra.}
\label{red}  
\begin{figure*}
\includegraphics[width=0.99\textwidth]{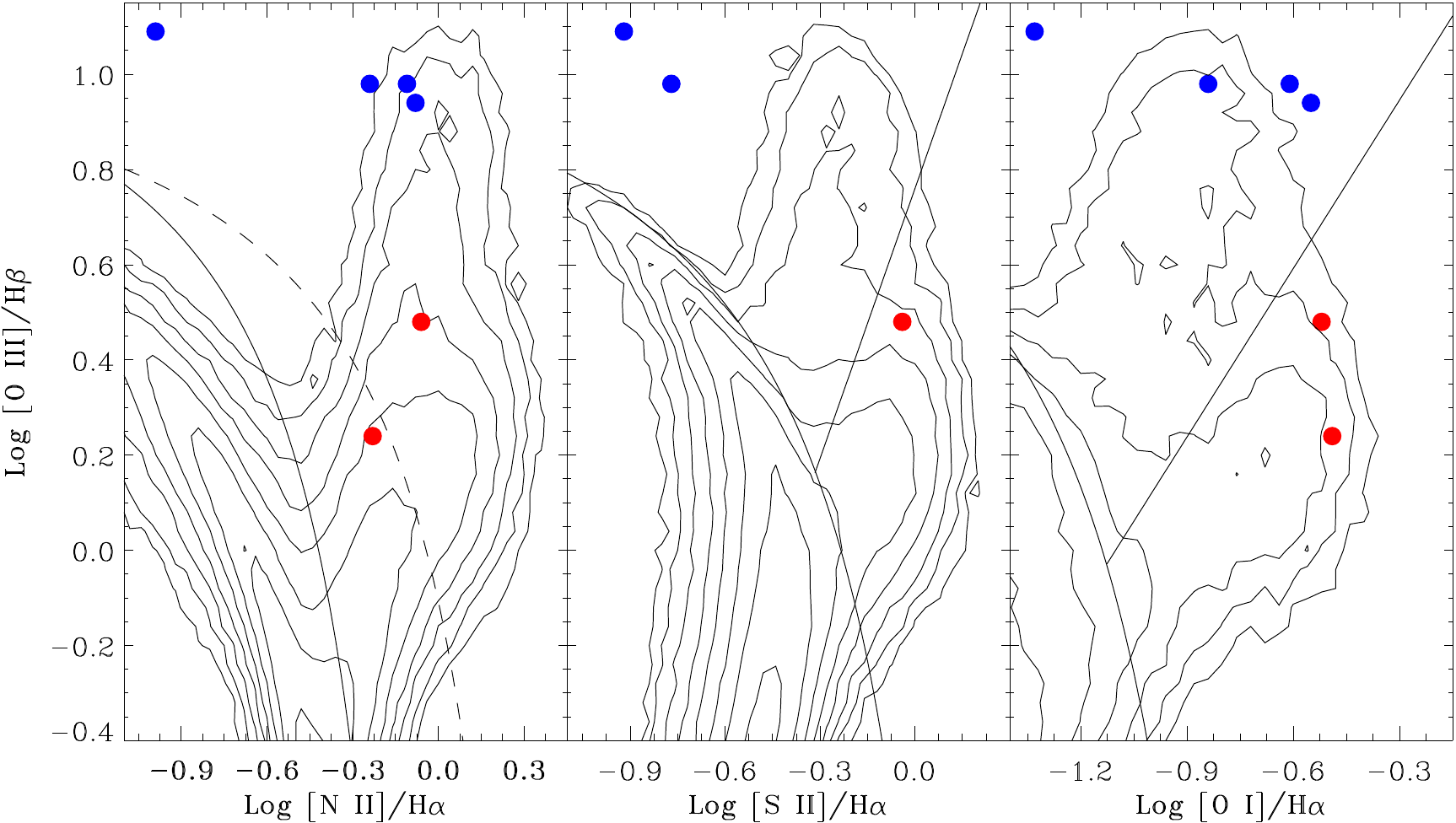}
\caption{Standard diagnostic diagrams for the six izRGs at the lowest redshift,
whose spectra include the [N~II]+\Ha\ complex. The blue (red) circles
  are the sources classified as HEGs (LEGs) based on the blue-DDs.}
\label{redDD}
\end{figure*}

We present the red portion of the nuclear spectra of the 26 izRGs
obtained from the MUSE observations focusing on the \Ha\ spectral
region.  The location of the six izRGs at the lowest redshift, whose
spectra includes the [N~II]+\Ha\ complex, is shown in
Fig. \ref{redDD}. The classification based on the standard DD is
consistent with that derived from the blue-DDs.

\begin{figure*}  
\includegraphics[width=0.45\textwidth]{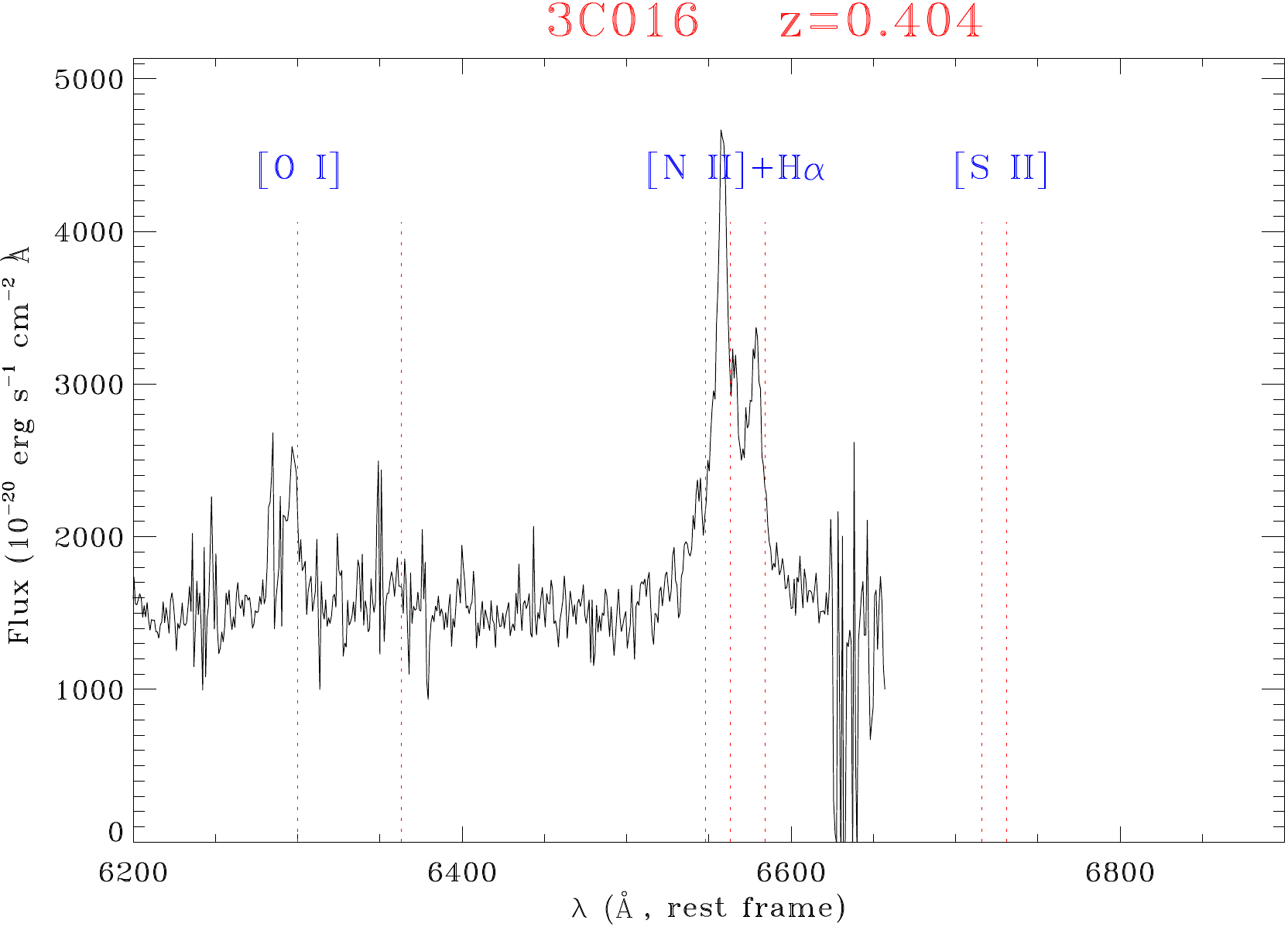}
\includegraphics[width=0.45\textwidth]{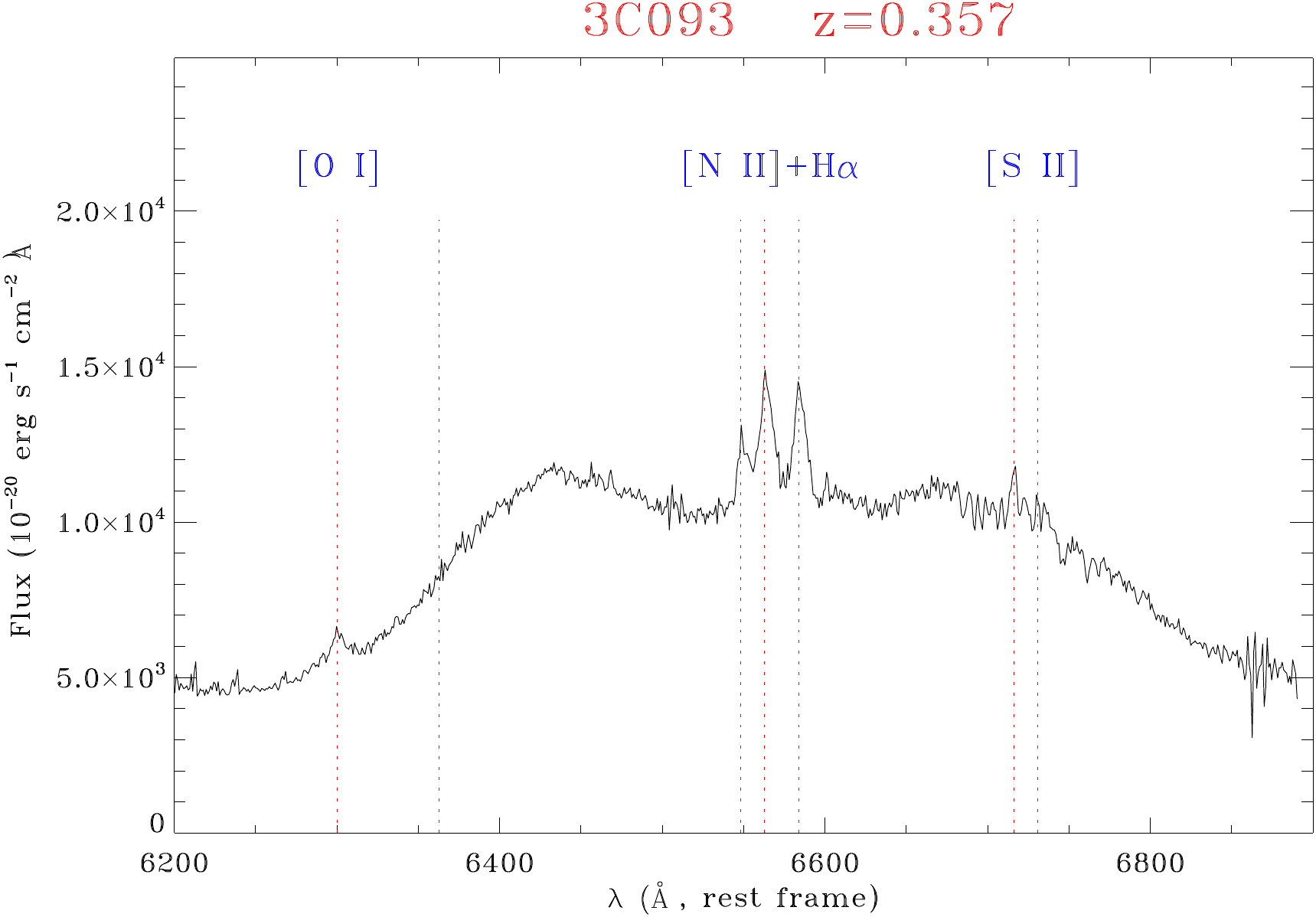}
\includegraphics[width=0.45\textwidth]{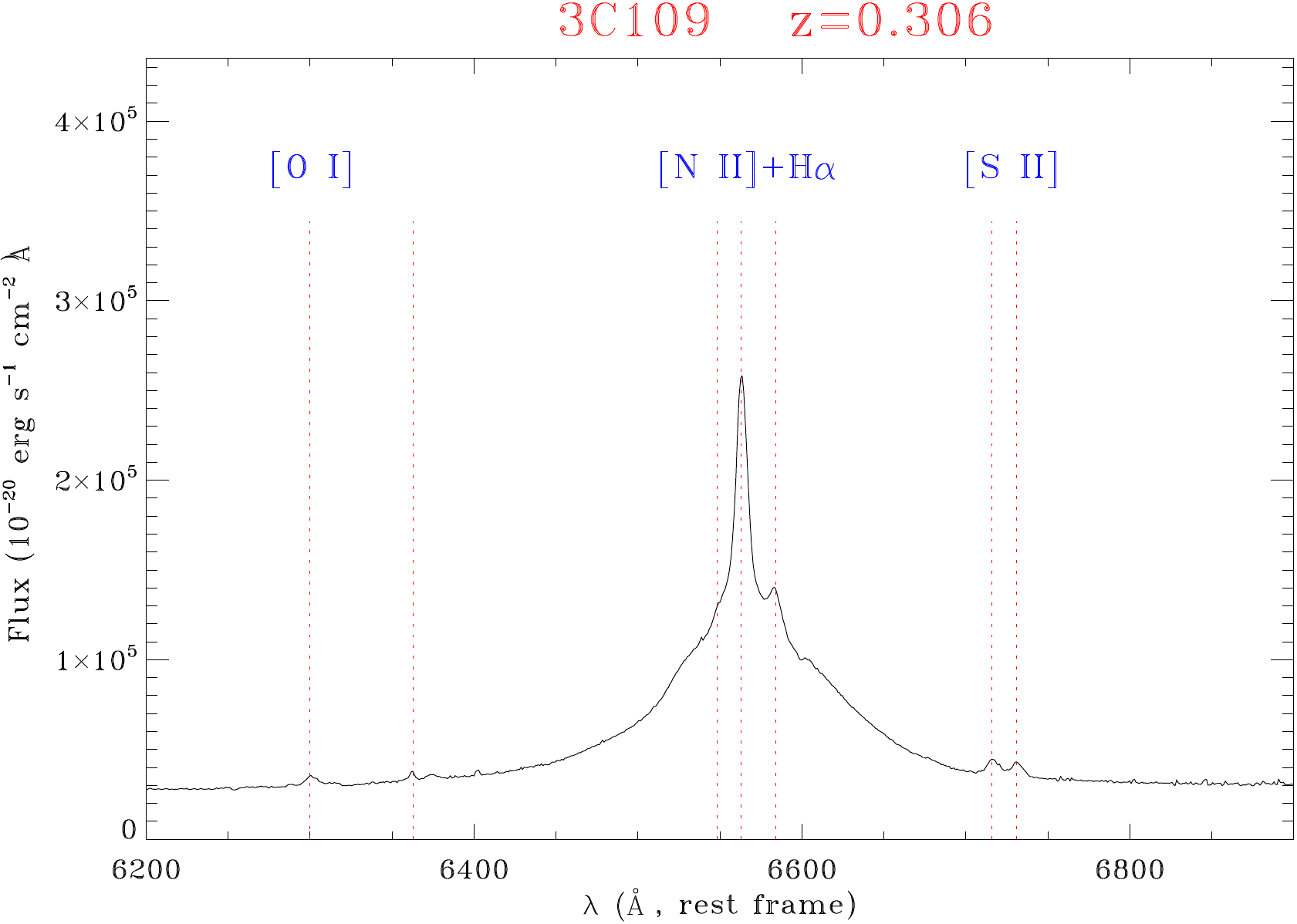}
\includegraphics[width=0.45\textwidth]{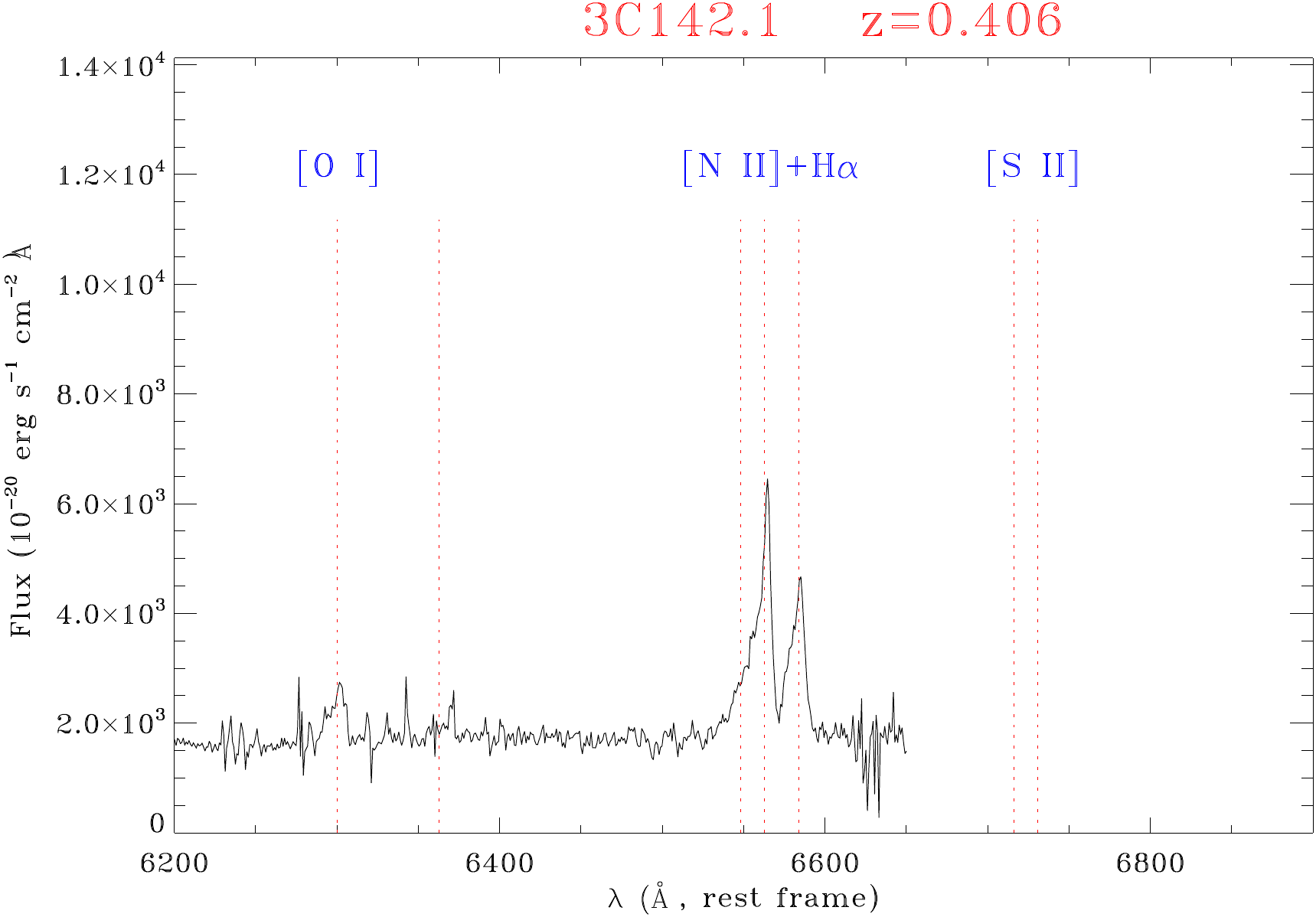}
\includegraphics[width=0.45\textwidth]{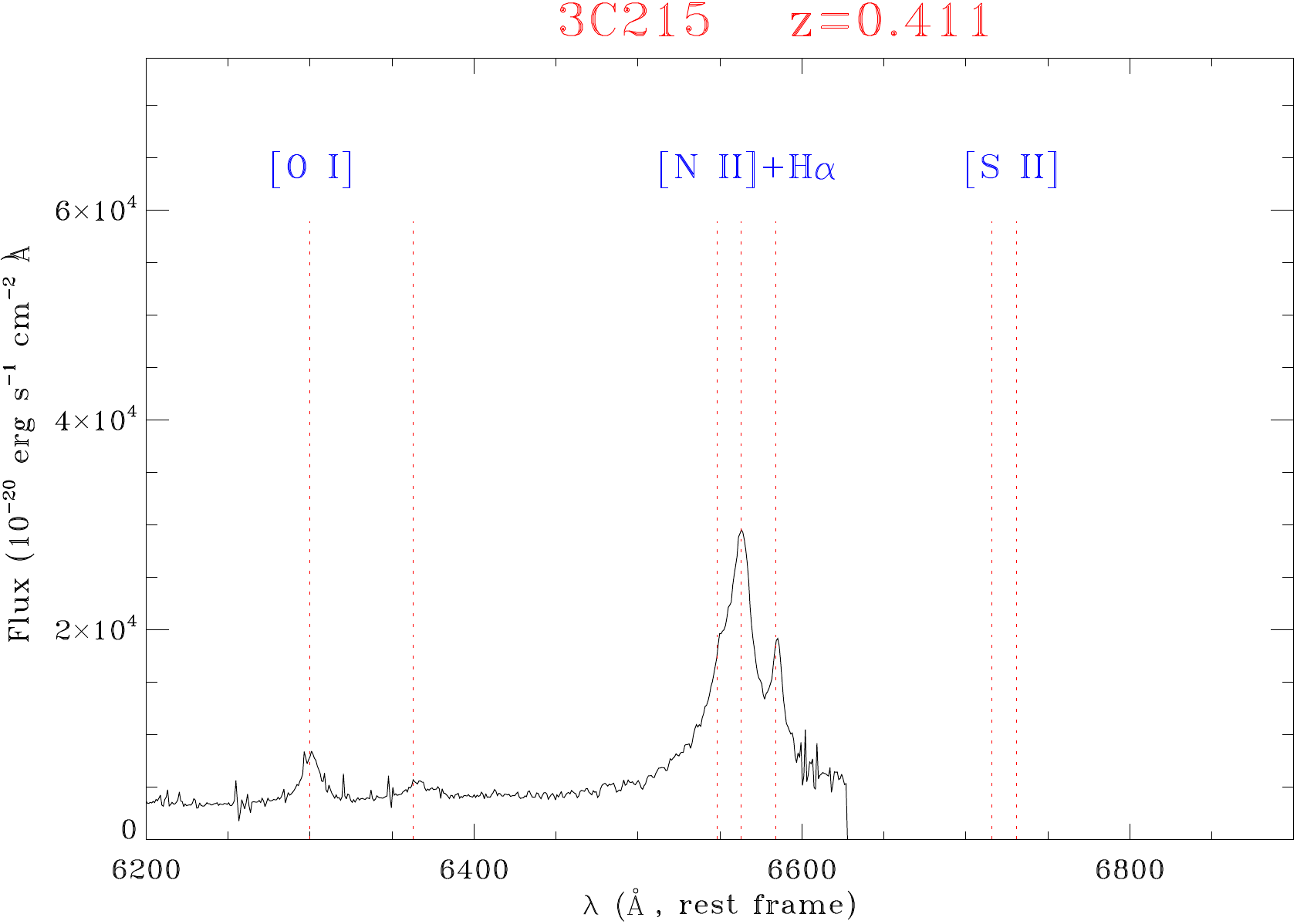}
\includegraphics[width=0.45\textwidth]{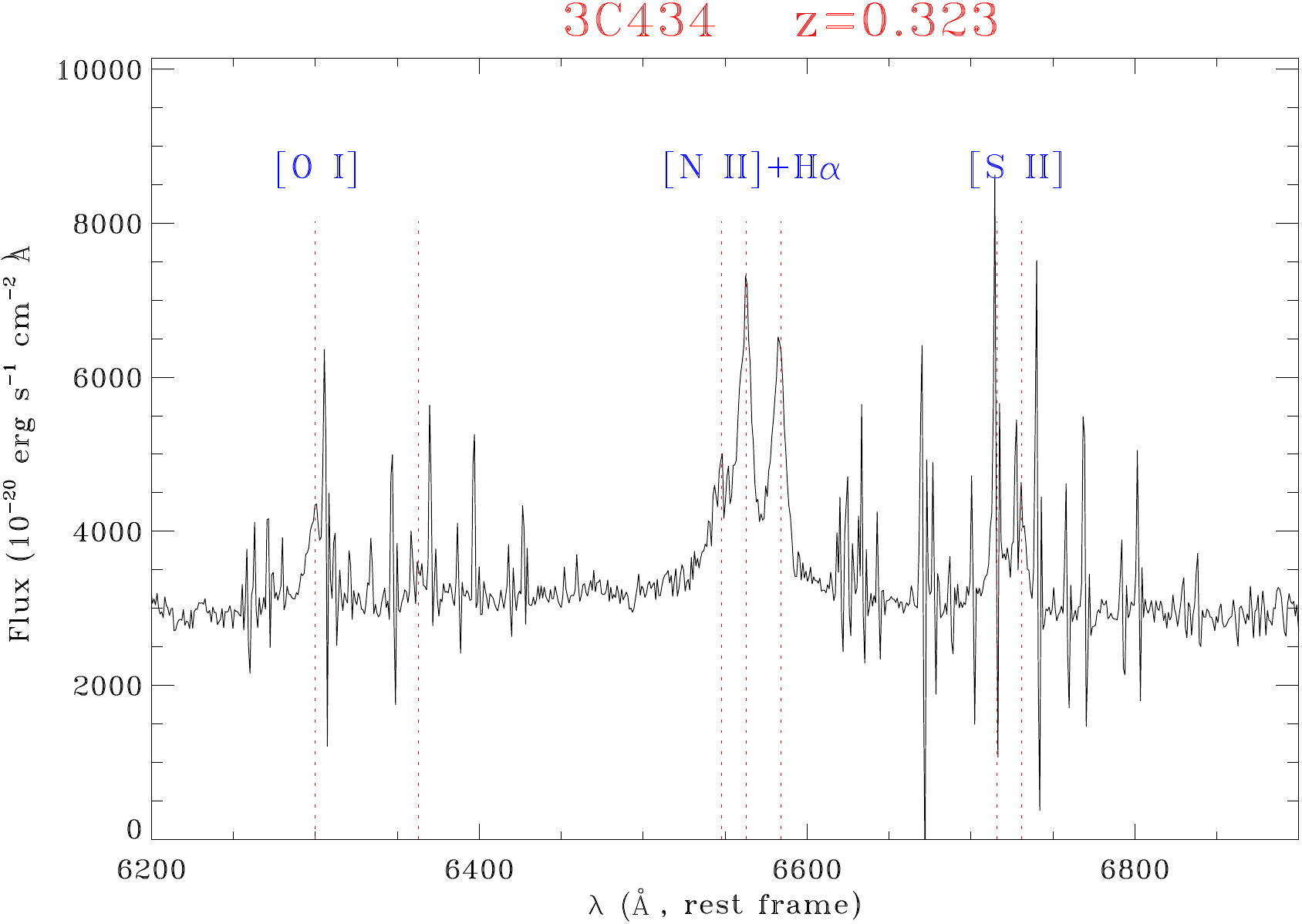}
\caption{Red portion of the nuclear spectra of the six lowest redshift
  izRGs.}
\label{redspectra}
\end{figure*}

\end{appendix}

\end{document}